\documentclass[hidelinks,journal=jctc,manuscript=article,
    11pt]{achemso}
\setkeys{acs}{doi = true}
\usepackage{xr}
\usepackage[table,xcdraw]{xcolor}
\usepackage{tikz}
\usetikzlibrary{positioning}
\usepackage[utf8]{inputenc}
\usepackage[T1]{fontenc}
\usepackage{filecontents,catchfile}
\usepackage{chemformula}
\usepackage{comment}
\usepackage{hyperref}
\usepackage{textgreek}
\usepackage{amsmath}
\usepackage{siunitx}
\usepackage{todonotes}
\usepackage{texshade}
\usepackage{lscape}
\makeatletter

\title{A Reversible Unwrapping Algorithm for Constant Pressure Molecular Dynamics Simulations}

\author{Martin Kulke}
\affiliation[MSU]
{MSU-DOE Plant Research Laboratory and Department of Biochemistry and Molecular Biology, Michigan State University, 612 Wilson Rd, East Lansing, MI 48824, United States of America.}
\author{Josh V Vermaas}
\affiliation[MSU]
{MSU-DOE Plant Research Laboratory and Department of Biochemistry and Molecular Biology, Michigan State University, 612 Wilson Rd, East Lansing, MI 48824, United States of America.}
\email{vermaasj@msu.edu}

\abbreviations{}

\keywords{Molecular Dynamics, Periodic Boundary Conditions, Unwrapping, NAMD, GROMACS}

\begin{document}
	
\begin{tocentry}
\includegraphics{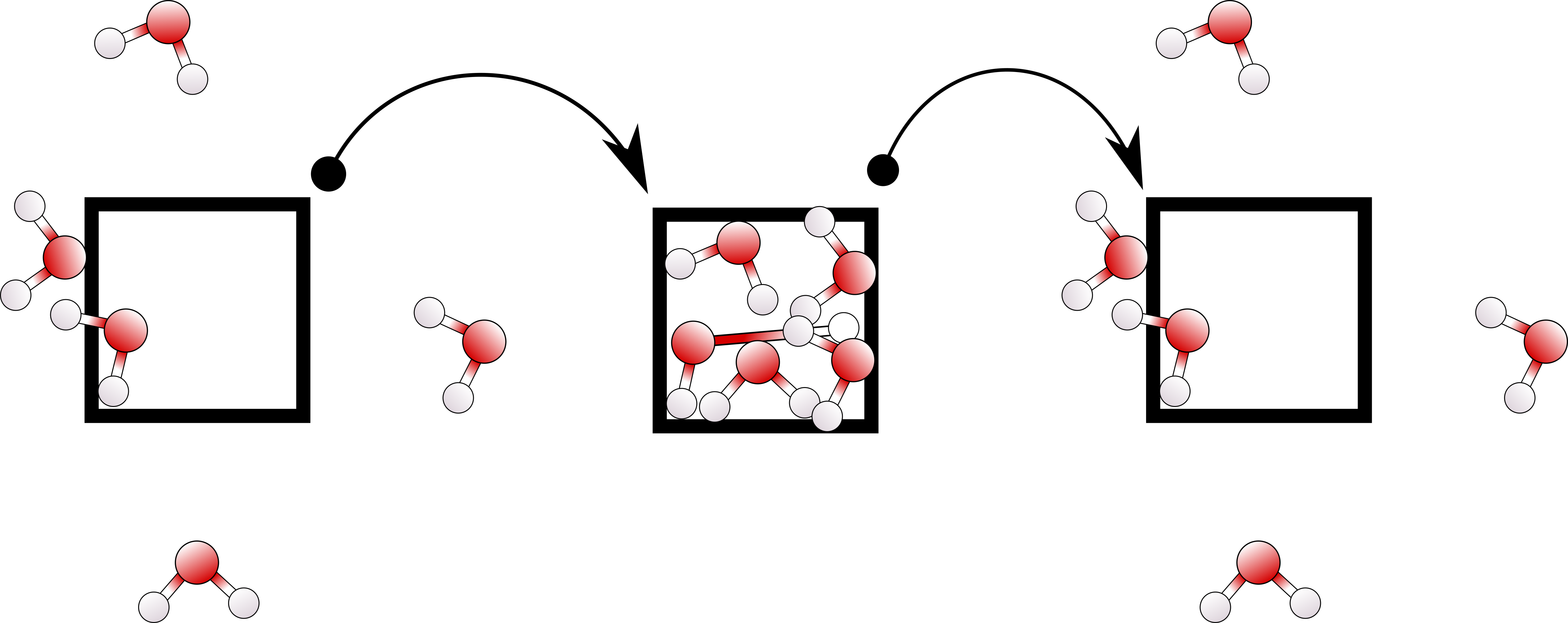}
\end{tocentry}

\begin{abstract}
Molecular simulation technologies have afforded researchers a unique look into the nanoscale interactions driving physical processes.
However, a limitation for molecular dynamics (MD) simulations is that they must be performed on finite-sized systems in order to map onto computational resources.
To minimize artifacts arising from finite-sized simulation systems, it is common practice for MD simulations to be performed with periodic boundary conditions (PBC).
However, in order to calculate specific physical properties, such as mean square displacements to calculate diffusion coefficients, continuous particle trajectories where the atomic movements are continuous and do not jump between cell faces are required.
In these cases, modifying atomic coordinates through unwrapping schemes are an essential post-processing tool to remove these jumps.
Here, two established trajectory unwrapping schemes are applied to \SI{1}{\micro\second} wrapped trajectories for a small water box.
The existing schemes can result in spurious diffusion coefficients, long bonds within unwrapped molecules, and inconsistent atomic coordinates when coordinates are rewrapped after unwrapping.
We determine that prior unwrapping schemes do not account for changing periodic box dimensions, and introduce an additional correction term to the existing displacement unwrapping scheme by von B\"ulow et al. to correct for these artifacts.
We also demonstrate that the resulting algorithm is a hybrid between the existing heuristic and displacement unwrapping schemes.
After treatment with this new unwrapping scheme, molecular geometries are correct even after long simulations.
In anticipation for longer molecular dynamics trajectories, we develop implementations for this new scheme in multiple PBC handling tools.
\end{abstract}

\section{Introduction}

Molecular dynamics simulations are a fantastic tool to study interactions and mechanisms in the nanoscale regime.
Applications range broadly from material studies \cite{Yang2019, Zepeda-Ruiz2017, Lau2018, Massobrio2015}, to disease mechanism \cite{Arantes2020, Hollingsworth2018,Perilla2017a, Casalino2020a, Durrant2020} and trying to understanding how life itself works \cite{Feig2017,Bock2018, Nawrocki2019, Kubo2020}.
Recent high profile applications include a recent study where the complete HIV-1 empty capsid was simulated revealing biological implications based on its physical properties.\cite{Perilla2017a}
Similarly, the simulation of SARS-CoV-2's spike protein provided a unique inside into the role of its glycosylations in evading detection.\cite{Casalino2020a}
The trend in the field is towards larger system sizes that cover longer simulation times, with the first reported billion atom biological systems having already been simulated\cite{Jung2019}.
From the simulation of a crowded bacterial cytoplasm\cite{Yu2016}, models for whole bacterial cells are under development\cite{Maritan2021}.

Inherent to the success of molecular dynamics, there are certain common methodological assumptions made in system construction.
Long range electrostatics is essential to accurately model these systems.
The truncation of long range electrostatic interactions with cutoffs leads to wrong structures for charged proteins\cite{Saito1994,Loncharich1989}, DNA deforms \cite{Norberg2000} and lipid bilayers have an incorrect area per lipid\cite{Patra2003}.
To calculate long-range electrostatics through particle mesh Ewald summations\cite{Ewald1921,Essmann1995}, and to mitigate edge effects for finite sized-systems, periodic boundary conditions (PBC) are imposed onto the molecular simulation systems.
With PBC, the volume is maintained during NVT simulations and it allows barostats\cite{Feller1995} to adjust the pressure in NPT simulations by changing the box volume.

A consequence of simulating with PBC is that atoms move across edges, exiting across one cell face and reentering the cell from the opposite face. %
Most molecular dynamics simulation packages keep track of the atomic positions only within the origin cell and output frames to trajectories accordingly.
Thus, most analysis is performed on a trajectory where the atomic coordinates are "wrapped" within a minimal unit cell.
However, some analysis requires that particle positions account for cell face crossing events, such that atomic coordinates are "unwrapped", as though PBC wrapping had not been applied.
A common use case for unwrapped trajectories is to calculate diffusion coefficients from atomic mean square displacements (MSD), to keep drifting multi-domain proteins in contact for visualization or analysis purposes, or to recenter lipids in membrane studies.

Since the advent of periodic boundary conditions in molecular simulation, simulation and analysis packages have determined the unwrapped coordinates at timestep i+1 ($x_{i+1}^{u}$) by tracking the number of periodic unit cell vectors that need to be added to account for crossing events across the unit cell boundary.
Concretely, the unwrapped coordinate is recalculated every step with the unwrapped coordinates of the previous step $x_{i}^{u}$, the current wrapped position $x_{i+1}^{w}$ and periodic unit cell dimensions $L_{i+1}$:

\begin{equation}
x_{i+1}^{u} = x_{i+1}^{w} - \left\lfloor \frac{x_{i+1}^{w}-x_{i}^{u}}{L_{i+1}}+\frac{1}{2}\right\rfloor L_{i+1}
\label{eq:heuristic}
\end{equation}

\noindent
We will refer to equation \ref{eq:heuristic} as the heuristic method, as this method is based on reversing the coordinate wrapping algorithm inside simulation engines.
The heuristic method has some appealing features, such as only requiring the previous unwrapped coordinates to be kept in memory while replacing the wrapped coordinates with the unwrapped coordinates in place.
Unwrapping using the heuristic method is also invertible, as typical wrapping algorithms will arrive at the exact number of box vectors needed to regenerate the wrapped coordinates, preserving relative atom positions for contact analysis or visualization.

However, \citet{VonBulow2020} demonstrated that the heuristic method yields incorrect diffusion constants during microsecond long simulations in NPT ensembles, especially for small boxes.
The spurious diffusion coefficients were hypothesized to arise from periodic cell dimension fluctuations due to the applied barostat\cite{VonBulow2020}.
When the unwrapped position becomes large relative to the unit cell dimensions, the number of unit cells to shift the wrapped position can vary depending on the instantaneous unit cell dimension at a given timestep.
Based on these findings, \citet{VonBulow2020} proposed an alternative unwrapping scheme, which we label displacement method in this study:

\begin{equation}
	x_{i+1}^{u} = x_{i}^{u} + (x_{i+1}^{w} - x_{i}^{w}) - \left\lfloor \frac{x_{i+1}^{w}-x_{i}^{w}}{L_{i+1}}+\frac{1}{2}\right\rfloor L_{i+1}
\label{eq:displacement}
\end{equation}

\noindent
Rather than counting the number of unit cell crossing events, the displacement scheme adds the wrapped displacement at every timestep, adjusting the displacement by the unit cell dimension, if the particle would otherwise travel more than half a box dimension between timesteps.
As the unwrapped trajectory is determined by a sequence of displacement vectors, and the displacement vectors are typically smaller than the unit cell length, the displacement scheme is more robust to unit cell fluctuations when in an NPT ensemble.
However, when applied to long simulation trajectories, the displacement scheme as written in Eq.~\ref{eq:displacement} has two noticeable flaws.
Under certain circumstances, the displacement scheme can generate long bonds indicative of a broken molecule.
Furthermore, the displacement scheme is not reversible with conventional wrapping schemes, and can subtly distort the molecular structure during analysis and visualization.

\section{Theory}

To understand, where the artifacts in the heuristic and displacement scheme originate from, we look at the formalism of these equations.
The discrete unwrapped atomic movement between two timepoints within a simulation can be described as a displacement $d$ between steps $i$ and $i+1$.

\begin{equation}
	d = x_{i+1}^{u} - x_i^u
	\label{eq:t_displacement}
\end{equation}

\noindent
To maintain reversibility between wrapping and unwrapping, unwrapped coordinates $x^u$ can be expressed as coordinates $x^w$ wrapped inside a periodic box and translated by an integer number of box dimensions $L$.
If this is not true, successive wrapping and unwrapping will not yield consistent numerical results.
As will be shown for the heuristic scheme, the number of periodic boxes translated between two frames $m,n \in \mathbb{Z}$ can vary significantly depending on the atom movement and the coordinate scaling magnitude from barostat pressure adjustments.

\begin{align}
	x_{i}^{u} &= x_{i}^{w} - m L_{i}
	\label{eq:t_mframe_boxes} \\
	x_{i+1}^{u} &= x_{i+1}^{w} - n L_{i+1}
	\label{eq:t_nframe_boxes}
\end{align}

\noindent
An expression for the periodic box number $n$ for frame $i+1$ is obtained by rearranging Eq. \ref{eq:t_nframe_boxes}, in which
the unwrapped coordinate $x_{i+1}^u$ of frame $i+1$ is an unknown quantity that is replaced with Eq. \ref{eq:t_displacement}.

\begin{equation}
	n = \frac {x_{i+1}^{w} - x_{i+1}^{u}} {L_{i+1}} = \frac {x_{i+1}^{w} - x_i^u - d} {L_{i+1}} 
	\label{eq:t_heuristic_1}
\end{equation}

\noindent
The true displacement $d$  is unknown \textit{a priori}.
However, we know that $n$ from Eq.~\ref{eq:t_heuristic_1} must be an integer.
Thus, assuming that the displacement is small enough relative to the box dimension that an atom will not move more than half a box vector between subsequent frames $| \frac{d}{L_{i+1}} |<\frac{1}{2}$, a floor function calculates $n$:

\begin{equation}
	n = \left\lfloor\frac{x_{i+1}^{w} - x_i^u}{L_{i+1}} - \frac{d}{L_{i+1}} + \frac{1}{2} \right\rfloor = \left\lfloor\frac{x_{i+1}^{w} - x_i^u}{L_{i+1}} + \frac{1}{2} \right\rfloor
	\label{eq:t_heuristic_final}
\end{equation}

\noindent
Inserting Eq. \ref{eq:t_heuristic_final} into Eq. \ref{eq:t_nframe_boxes} yields the heuristic unwrapping equation (Eq. \ref{eq:heuristic}).

The displacement method (Eq.~\ref{eq:displacement}) can be derived by replacing the unwrapped coordinate expressions in Eq. \ref{eq:t_displacement} with their definitions from Eq. \ref{eq:t_mframe_boxes} and \ref{eq:t_nframe_boxes} giving rise to two box size dependent terms.

\begin{equation}
	\begin{split}
	x_{i+1}^{u} - x_i^u &= x_{i+1}^{w} - n L_{i+1} - ( x_{i}^{w} - m L_{i} ) \\
	                    &= (x_{i+1}^{w} - x_{i}^{w}) - n L_{i+1} + m L_{i} { \color{red} + m\left( L_{i+1} - L_{i+1}\right)} \\
	                    &= (x_{i+1}^{w} - x_{i}^{w}) - (n-m) L_{i+1} - m (L_{i+1}-L_{i})
	\label{eq:t_displacement_1}
	\end{split}
\end{equation}

\noindent
From Eq. ~\ref{eq:t_displacement_1}, we see a dependence on the wrapped displacement $x_{i+1}^{w} - x_{i}^{w}$ similar to the displacement algorithm from \citet{VonBulow2020} (Eq.~\ref{eq:displacement}).
However, we also see that the box dimension L at both time $i$ and $i+1$ make an appearance in the equation, which we simplify by adding zero through the red term in Eq. ~\ref{eq:t_displacement_1}.
When written in this way, we see that in addition to the wrapped displacement, there is a dependence on the current box dimension $L_{i+1}$, as well as on the change in box dimension between frames, $L_{i+1}-L_{i}$.
We find that there are two critical integers that need to be calculated, $n-m$, and $m$ itself.
$m$ is straightforward to calculate from its definition in Eq.~\ref{eq:t_mframe_boxes}.
\begin{equation}
	m = \frac{x_{i}^{w} - x_{i}^{u}}{L_{i}} = \left\lfloor\frac{x_i^w - x_i^u}{L_i} + \frac{1}{2}\right\rfloor
	\label{eq:t_displacement_2}
\end{equation}
\noindent
The floor function within Eq.~\ref{eq:t_displacement_2}  ensures that $m$ is an integer value even after potential floating point arithmetic errors.
Similarly, we can compute $n-m$ directly from its definition, and arrive at an analogous equation to Eq.~\ref{eq:t_displacement_2}, where $n-m$ is defined in terms of the unknown $x_{i+1}^u$.
Alternatively, we rearrange Eq.~\ref{eq:t_displacement_1} and insert the definition for the displacement $d$ to yield:
\small
\begin{equation}
	\begin{split}
		n - m &=  \left\lfloor\frac{x_{i+1}^{w} - x_{i}^{w}}{L_{i+1}} + m\frac{ L_{i}}{L_{i+1}} - \frac{d}{L_{i+1}} + \frac{1}{2} \right\rfloor - m \\
		&= \left\lfloor\frac{x_{i+1}^{w} - x_{i}^{w}}{L_{i+1}} + m\left(\frac{ L_{i}}{L_{i+1}} - 1\right) - \frac{d}{L_{i+1}} + \frac{1}{2} \right\rfloor \\
		&= \left\lfloor\frac{x_{i+1}^{w} - x_{i}^{w}}{L_{i+1}} + m\left(\frac{ L_{i}-L_{i+1}}{L_{i+1}}\right) - \frac{d}{L_{i+1}} + \frac{1}{2} \right\rfloor
		\label{eq:t_displacement_3}
	\end{split}
\end{equation}
\normalsize
\noindent
If we again assume that $d$ is sufficiently small such that $|  m\left(\frac{ L_{i}-L_{i+1}}{L_{i+1}}\right) - \frac{d}{L_{i+1}} |<\frac{1}{2}$, thereby flooring to an integer, the definition for $n-m$ becomes:

\begin{equation}
	n-m = \left\lfloor\frac{x_{i+1}^{w} - x_{i}^{w}}{L_{i+1}} + \frac{1}{2} \right\rfloor
	\label{eq:t_displacement_3a}
\end{equation}

\noindent
Eq.~\ref{eq:t_displacement_3a} matches the structure for the box correction term from Eq.~\ref{eq:displacement}.
At first glance, it appears that $|  m\left(\frac{ L_{i}-L_{i+1}}{L_{i+1}}\right) - \frac{d}{L_{i+1}} |$ is generally larger then the heuristic scheme equivalent $| \frac{d}{L_{i+1}} |$.
We will demonstrate from trajectory data that this is not the case and, in fact, the term is smaller.
Inserting Eq.~\ref{eq:t_displacement_2} and Eq.~\ref{eq:t_displacement_3a} into Eq.~\ref{eq:t_displacement_1} yields the following unwrapping equation:

\begin{equation}
	\begin{split}
		x_{i+1}^{u} &= x_{i}^{u} + (x_{i+1}^{w} - x_{i}^{w}) - (n-m) L_{i+1} - m (L_{i+1}-L_{i}) \\
				    &= x_{i}^{u} + (x_{i+1}^{w} - x_{i}^{w}) - \left\lfloor\frac{x_{i+1}^{w}-x_{i}^{w}}{L_{i+1}}+\frac{1}{2}\right\rfloor L_{i+1} - \left\lfloor\frac{x_{i}^{w}-x_{i}^{u}}{L_{i}}+\frac{1}{2}\right\rfloor(L_{i+1}-L_{i})
		\label{eq:correction}
	\end{split}
\end{equation}

In this study, we examine the conditions during a standard NPT simulation where Eq.~\ref{eq:heuristic} and \ref{eq:displacement} fail, and test the alternative unwrapping scheme from Eq.~\ref{eq:correction} that corrects Eq.~\ref{eq:displacement} by explicitly accounting for fluctuating unit cell dimensions.
The corrected hybrid scheme retains the best features from previous unwrapping methods.
The hybrid scheme is based on the same assumption as the heuristic method, in that the unwrapped and wrapped coordinates are related by an integer number of box vectors, and is thus invertible.
However, the hybrid scheme also has similarities in functional form to the displacement scheme, and therefore also correctly captures diffusion like the displacement scheme does.
The benefit to this hybrid approach is that there are fewer structural deformations, allowing for longer timesteps before structural artifacts emerge.

\section{Methods}

To thoroughly explore the three different unwrapping schemes, we generate multiple wrapped trajectories through two molecular dynamics engines in both constant volume (NVT) and constant pressure (NPT) ensembles.
After unwrapping, we evaluate the trajectories by measuring diffusion coefficients and  bond lengths within all water molecules.
To measure the reversibility for the unwrapping procedure, as might be needed for visualizing and analyzing multidomain protein simulations, we subsequently also wrap the coordinates once more and compare the resulting geometry with the initial coordinates.
This provides three metrics with which we can evaluate the different unwrapping schemes for their suitability in general use.

\subsection{Molecular dynamics simulations}

510 TIP3P water molecules\cite{Jorgensen1983,Neria1996} were placed in a \SI[product-units = single]{2.5 x 2.5 x 2.5}{\nano\meter\cubed} simulation cell via the \texttt{gmx solvate} tool of the GROMACS 5.1 software package \cite{Abraham2015}.
The water box was then minimized and equilibrated for \SI{70}{\pico\second} in a NPT ensemble at \SI{300}{\kelvin} and \SI{1}{atm} prior to production simulations.
The equilibrated water box starting point was simulated in both an NPT and NVT ensemble for \SI{1}{\micro\second} with wrapped trajectory snapshots saved every  \SI{20}{\femto\second}.
The frequent trajectory writes allowed for diffusion analysis to be performed using different lag times.
Both ensemble simulations were performed with the software packages GROMACS 2020\cite{Abraham2015,Pall2020} and NAMD 3.0a8\cite{Phillips2020}.
As an additional point of comparison, NAMD has the feature to write out unwrapped trajectory snapshots, which allowed us to simulate a NPT reference trajectory for our analysis.

In the GROMACS simulations, the equation of motion was integrated every \SI{2}{fs} by a leap-frog integrator.
Distances within water molecules are restrained with the SETTLE algorithm.\cite{Miyamoto1992}
Intermolecular interactions are considered within \SI{1}{nm} for Coulomb and van der Waals forces.
The low cutoff values allowed the use of small box sizes even with large volume fluctuations in the NPT ensemble.  
Long range electrostatic interactions are calculated with the fast smooth particle-mesh Ewald (PME) summation method, using a \SI{0.12}{nm} grid spacing.\cite{Ewald1921,Essmann1995}
The pair list is generated within \SI{1.1}{nm}.
Pressure and temperature are controlled by a Berendsen barostat\cite{Berendsen1984} to \SI{1}{atm} and modified v-rescale thermostat\cite{Bussi2007} to \SI{300}{\kelvin}, respectively.

NAMD uses the velocity verlet integrator\cite{Allen1987} to advance atomic position and velocities every \SI{2}{fs}, using the RATTLE algorithm\cite{Miyamoto1992} to restrain bond lengths to hydrogen.
Interaction pairs are generated every 10 steps between atom distances up to \SI{12}{\angstrom}.
Short-range electrostatic and van der Waals interactions use an identical \SI{10}{\angstrom} cutoff as our GROMACS trajectories.
Interaction forces are scaled down to zero with a switching function starting at \SI{9}{\angstrom}.
Long-range electrostatics are considered with PME using a grid spacing of \SI{1}{\angstrom}.\cite{Ewald1921,Essmann1995}
A Langevin piston\cite{Feller1995} and Langevin thermostat\cite{Grest1986} were used to adjust the temperature to \SI{300}{\kelvin} and the pressure to \SI{1}{atm}.

\subsection{Analysis}

The heuristic method had been previously implemented in \texttt{pbctools} 3.0 and \texttt{fastpbc} 1.0, while \texttt{qwrap} 1.4\cite{Henin2020} provided an implementation for the displacement method accessible from within VMD 1.9.4a48.\cite{Humphrey1996}
Our initial implementation for Eq.~\ref{eq:correction} was developed for \texttt{pbctools} 3.0.
We later implemented Eq.~\ref{eq:correction} into \texttt{fastpbc} and \texttt{qwrap} to improve execution time for the unwrapping procedure.
All implementations are provided in github repositories (for \href{https://github.com/kulkem/pbctools.git}{pbctools} and \href{https://github.com/kulkem/qwrap.git}{qwrap}\cite{Henin2020}), or have already been integrated into the VMD source tree (for fastpbc).
Together with supporting libraries such as numpy\cite{Harris2020}, sklearn\cite{Pedregosa2011}, and matplotlib\cite{Hunter2007}, we leveraged the VMD python API to gather, store, and visualize bond-length distributions, compute mean squared displacements, and compare coordinates between the original trajectories and the same trajectories after applying both wrapping and unwrapping operations.
Mean square displacements over time were used to determine diffusion coefficients from the Einstein relation\cite{Einstein1905}:
\begin{equation}
	D=\frac{\Delta(MSD)}{6\Delta t}
	\label{eq:diffusion}
\end{equation}
and subsequently using the sklearn\cite{Pedregosa2011} to determine a linear fit between MSD and time.
The source code to build these analyses is available on \href{https://www.researchgate.net/publication/356424261_Data_set_for_Investigating_Reversibility_in_Unwrapping_Schemes_for_Molecular_Dynamics_Simulations}{ResearchGate}.

\begin{figure*}[ht]
	\centering
	\includegraphics[width=\linewidth]{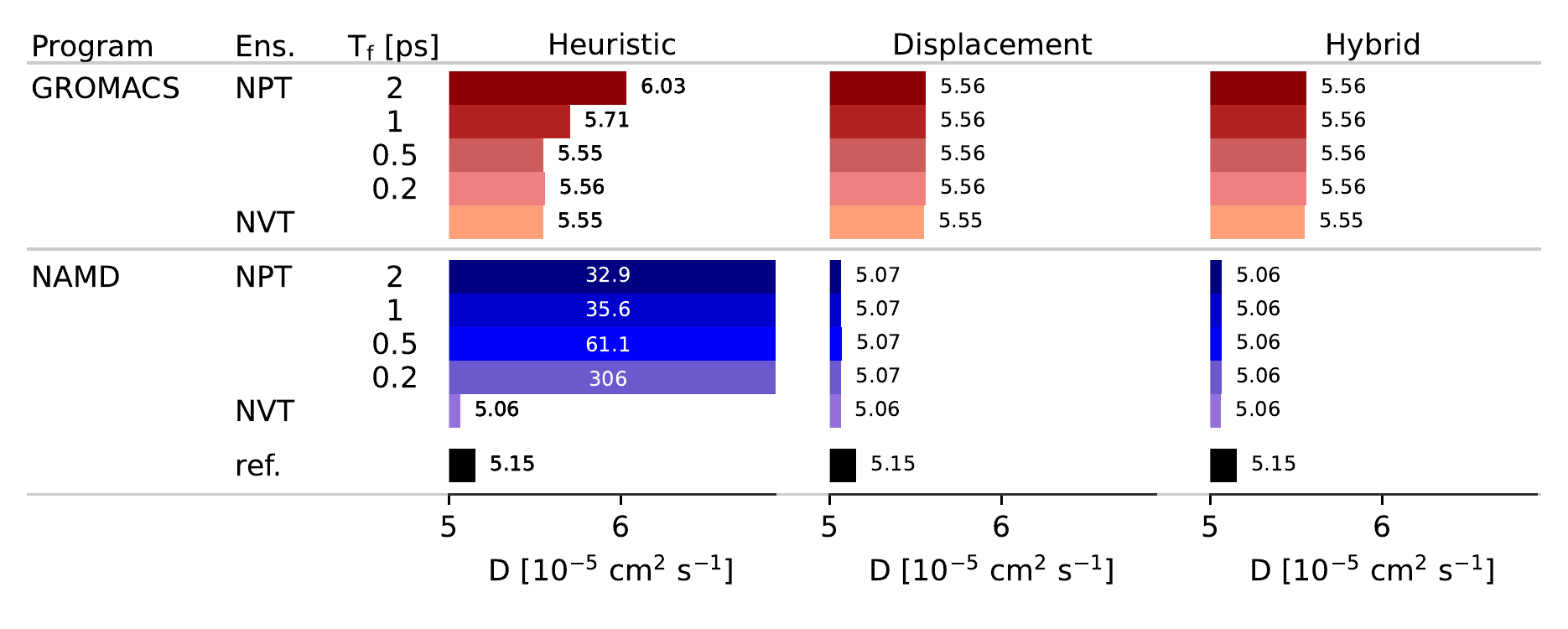}
	\caption{Diffusion coefficients $D$ for GROMACS and NAMD with different trajectory frequencies $T_f$.
		Reference values are calculated from the NAMD NPT ensemble simulation with unwrapped trajectory snapshots.}
	\label{fig:msd}
\end{figure*}

\section{Results and Discussion}

\subsection{Diffusion coefficients}

The motivation for \citet{VonBulow2020} to propose alternative unwrapping schemes focused on how the heuristic scheme overestimates diffusion coefficients for simulations conducted in NPT ensembles.
Analysis for our own independent simulations find similar results (\autoref{fig:msd}, S1-31), where the heuristic scheme can overstate the diffusion coefficient by up to a factor of 60.
The unit cell size changes caused by the barostat within a NPT ensemble are clearly the underlying cause for this issue, as the diffusion estimates agree between unwrapping methods for NVT-generated trajectories.

One hypothesis proposed by \citet{VonBulow2020} is that the displacement and heuristic methods would agree once the trajectory frequency analysis rates trend towards zero.
From \autoref{fig:msd}, we see that this is true only for GROMACS trajectories, and that instead NAMD trajectories have increased apparent diffusion after unwrapping with less time between frames.
This counter-intuitive finding appears to be related to the variation in the unit cell dimensions within a NAMD trajectory, which vary more quickly than what was observed for a GROMACS trajectory (\autoref{fig:si_fluctuations}).
The heuristic scheme only considers the box dimension for the current timestep, ignoring the previous timestep entirely.
Faster variation in the box dimension would lead to more instances where the integer $n$ from Eq.~\ref{eq:t_heuristic_final} would not match from one timestep to the next, increasing the error rate.

The displacement and hybrid methods result in the same diffusion coefficients, despite their different formalisms.
The key difference is that the hybrid method adds a term dependent on the change in periodic box dimensions, $L_{i+1}-L_{i}$.
If the box dimensions are near equilibrium, and are not persistently growing or shrinking over time, that would imply that  $\left<L_{i+1}-L_{i}\right>=0$, and the additional term in the hybrid scheme would contribute nothing to the overall diffusion behavior.

While the simulation conditions between NAMD and GROMACS trajectories are mostly similar, \autoref{fig:msd} does highlight some minor differences in the observed diffusion coefficients.
The most obvious difference between the simulations are the barostats and thermostats, which likely account for a majority of the differences seen in \autoref{fig:msd}.
Other implementation choices, such as the subtly different choices for physical constants, change the energies and forces computed at each timestep.\cite{Vermaas2016}
It is less clear why the diffusion coefficients for the unwrapped and wrapped NAMD trajectories differ.
We suspect that the force-position dot product within the virial calculation that controls the pressure is sensitive to the effective change in numerical precision between the wrapped and unwrapped coordinates, as the larger magnitude for the unwrapped coordinates reduces the precision for arithmetic operations.
Testing this hypothesis explicitly goes beyond the scope of this work.

\begin{figure*}[!ht]
	\centering
	\includegraphics[width=0.8\linewidth]{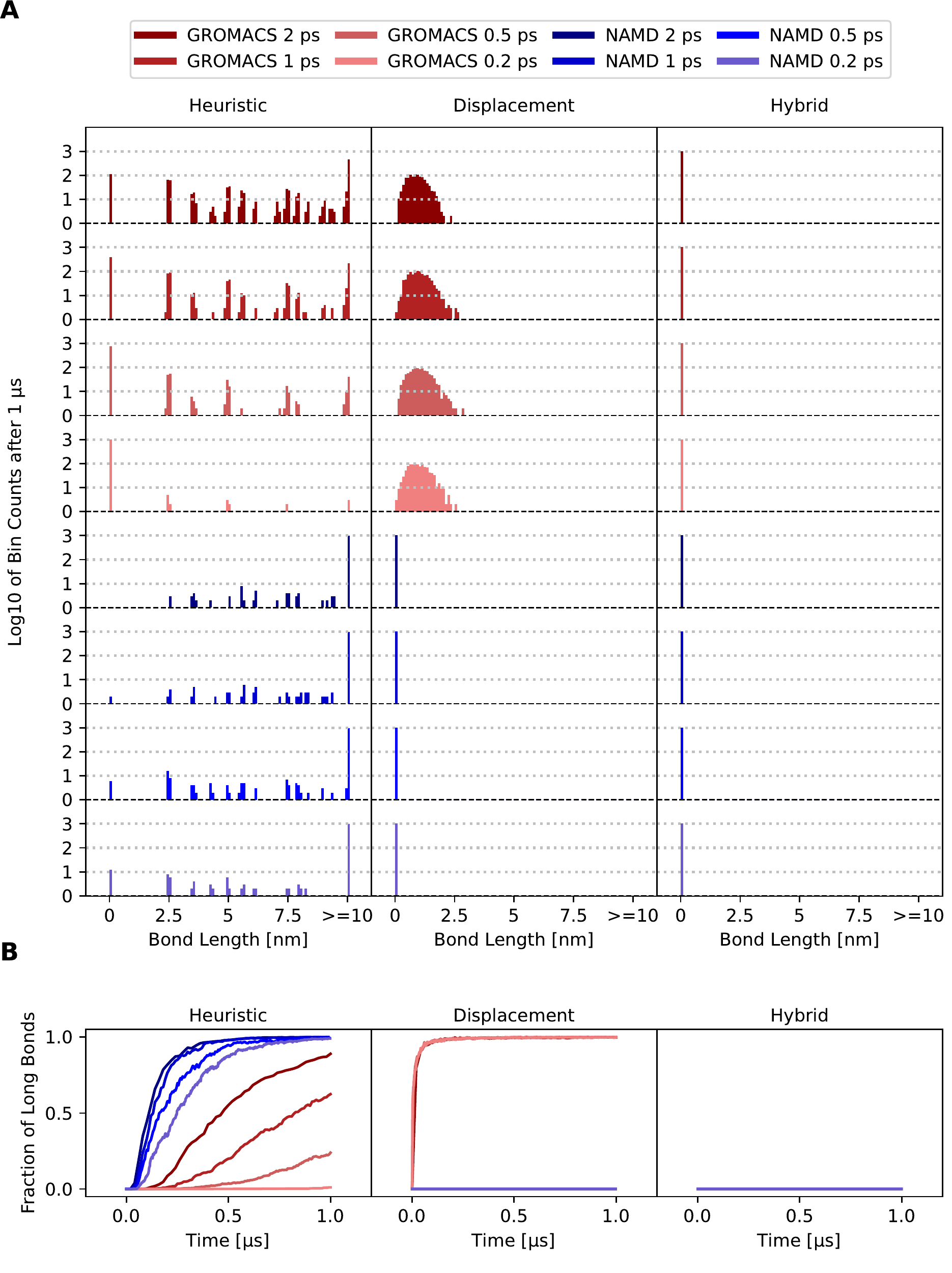}
	\caption{(A) Histograms of bond lengths after \SI{1}{\micro\second} for different trajectory frequencies in NAMD (red) and GROMACS (blue) using the three unwrapping implementations defined by Equations~\ref{eq:heuristic}-\ref{eq:correction}.
		Each row within the figure highlights a different combination of simulation engine and trajectory frequency, with the distribution colored according to the legend above.
		Each bin starts with a value of one to prevent the undefined logarithm of zero.
		     (B) Fraction of bond lengths over 0.95 nm during the simulation.
		 }
	\label{fig:bonds}
\end{figure*}

\subsection{Bond length distributions}

A correctly implemented unwrapping scheme should move the atomic positions only by multiples of the box dimension for a given frame.
Consequently, bond lengths between atoms should not change in an unwrapped trajectory, as a molecule should move together as one unit during simulations.
To verify correctness for the existing heuristic (Eq.~\ref{eq:heuristic}) and displacement (Eq.~\ref{eq:diffusion}) schemes, we calculated the bond lengths over long trajectories after unwrapping (\autoref{fig:bonds}), and compared them against our new hybrid scheme from Eq.~\ref{eq:correction}.
Since our system is entirely composed of water, bond lengths are expected to be uniformly slightly less than  \SI{1}{\AA}.
However, for both the heuristic and displacement schemes, we observe longer bond lengths on the nanometer length scale, indicative of stretched and unphysical water molecule geometries (\autoref{fig:bonds}).

The mechanism for generating the stretched bonds for the heuristic and displacement schemes are unique between the methods.
As seen in \autoref{fig:bonds}A, the spurious bond lengths for the heuristic method are primarily quantized in increments of \SI{2.5}{\nano\meter}, the PBC dimensions.
Diagonal long bonds are also present, as evidenced by 3.5 and \SI{4.3}{\nano\meter} bond lengths.
It has been hypothesized that the discrete bonds are the result of atomic motion beyond half of a unit cell between trajectory frames.
However, atoms have been measured to move around \SI{2.5}{\angstrom} on average between frames for the \SI{2}{\pico\second} trajectory frequency (\autoref{fig:si_disphist}).
While we cannot exclude that atoms moved more then half the box size in our simulations between trajectory writes, long bonds occur too frequently and only after a significant long simulated trajectory, suggesting that an alternative mechanism is at play.

\begin{figure}[ht]
	\centering
	\includegraphics[width=0.45\linewidth]{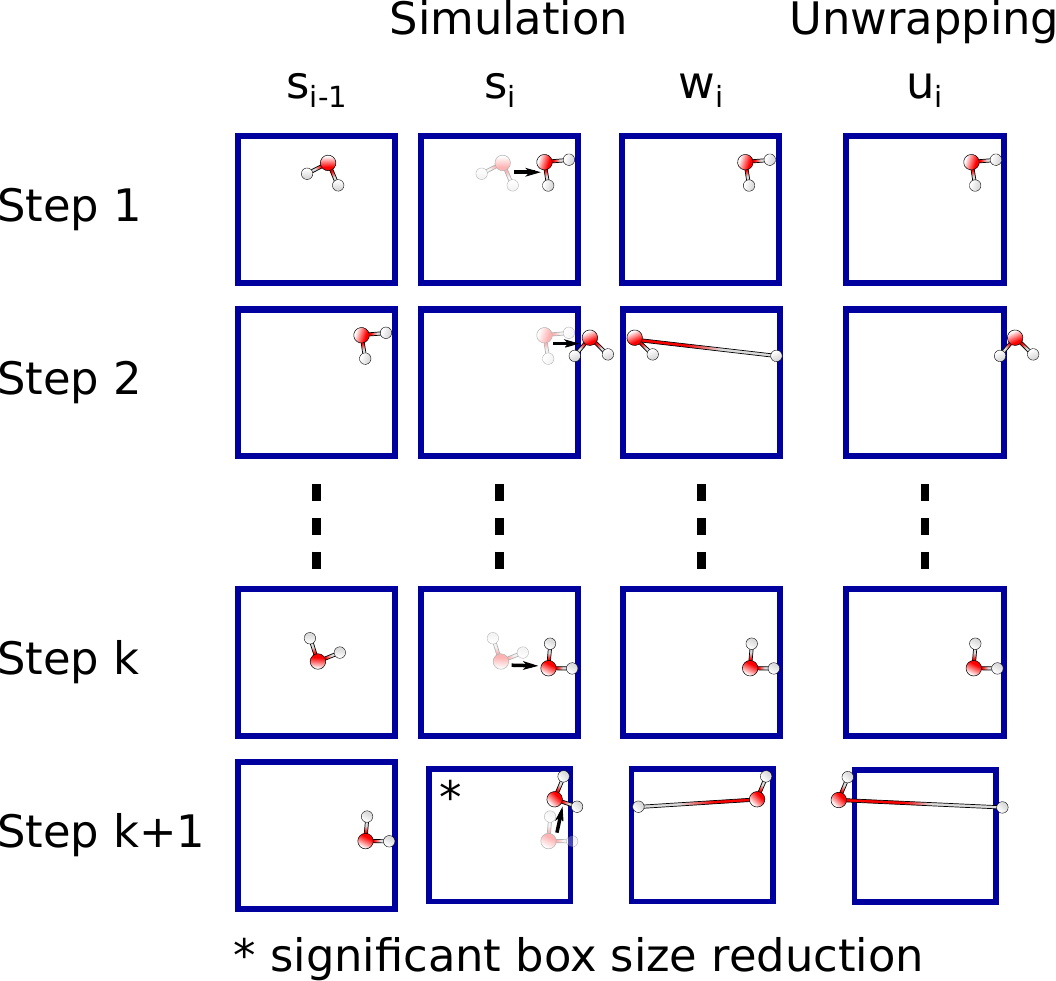}
	\caption{Schematic of the heuristic method showing an unwrapping error occurring at step $k+1$ after a significant box size reduction.
		During the simulation atoms move between time steps $s_{i-1}$ and $s_i$. 
		Afterwards the positions $w_i$ are wrapped.
		In the post-processing analysis the coordinates $u_i$ are unwrapped.}
	\label{fig:schemata}
\end{figure}

\begin{figure}[ht!]
	\centering
	\includegraphics[width=0.45\linewidth]{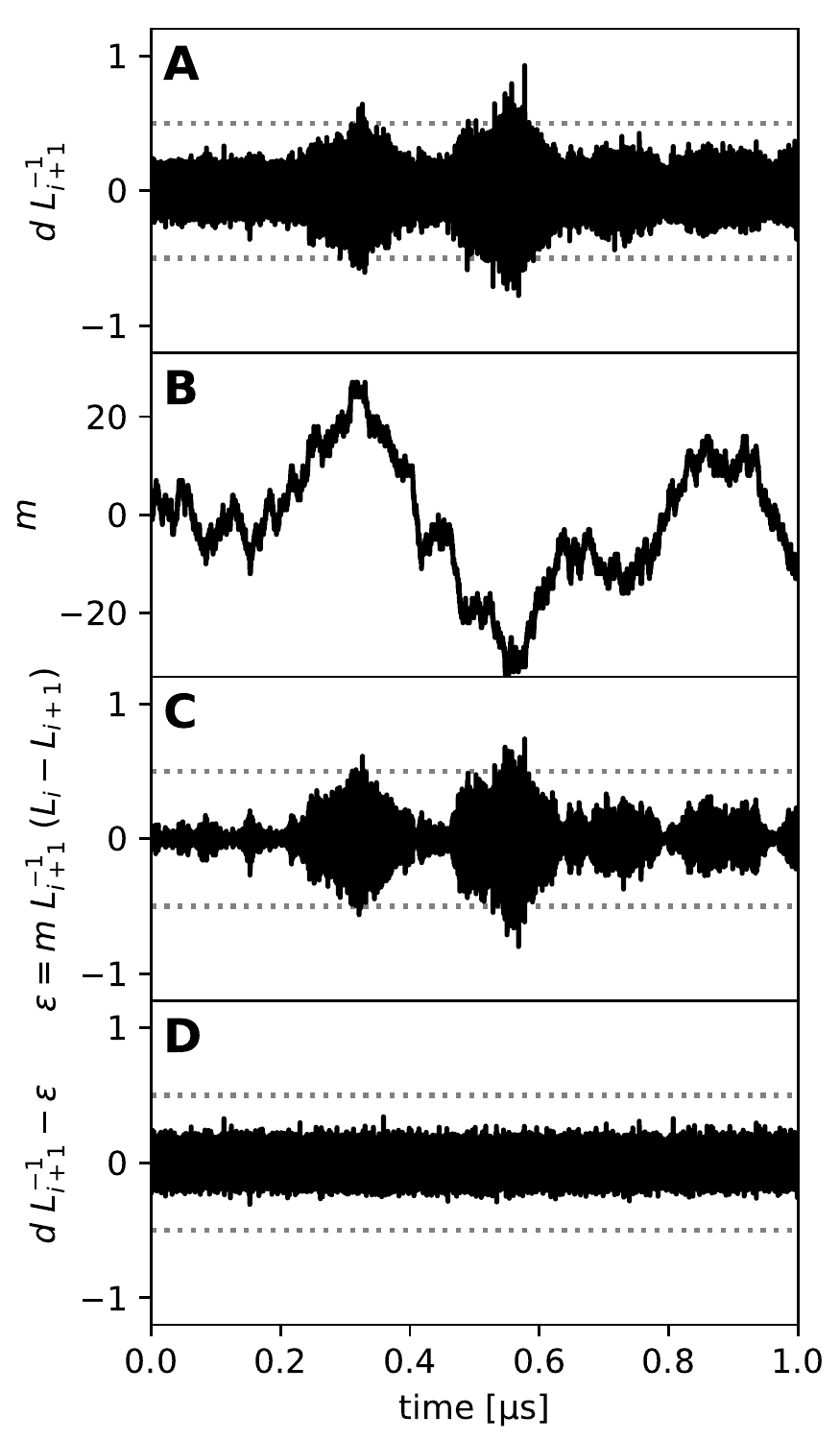}
	\caption{The (A) heuristic $| \frac{d}{L_{i+1}} |<\frac{1}{2}$ and (D) displacement $| m\left(\frac{ L_{i}-L_{i+1}}{L_{i+1}}\right) - \frac{d}{L_{i+1}} |<\frac{1}{2}$ unwrapping assumptions for one oxygen atom's x-coordinate.
		The terms are calculated from a NAMD trajectory without wrapping.
		The dotted lines indicate $\pm\frac{1}{2}$, with values above the upper or below the lower boundary violating the respective unwrapping assumptions leading to an unwrapping error.
		(B) The traversed simulation boxes Eq.~\ref{eq:t_displacement_2} of the atom.
		(C) As (A), but for the $m$ dependent term in the displacement unwrapping assumption.}
	\label{fig:floor_error}
\end{figure}

To demonstrate this mechanism, it is informative to have a concrete example.
At a given time well into the simulation trajectory, the unit cell dimension is \SI{25.24}{\angstrom}.
Two atoms $a_1$ and $a_2$ on the same molecule are at wrapped x-coordinates $x_1^w=9.12\,\text{\normalfont\AA}$ and $x_2^w=10.06\,\text{\normalfont\AA}$, respectively, connected by a standard \SI{1}{\angstrom} bond.
The unwrapped coordinates $x_1^u=463.43\,\text{\normalfont\AA}$ and $x_2^u=464.38\,\text{\normalfont\AA}$ are much larger, as the atoms have already traversed 18 simulation boxes.
In the next time step, both atoms move backwards and exchange relative orientations to $x_1^w=6.43\,\text{\normalfont\AA}$ and $x_2^w=5.46\,\text{\normalfont\AA}$.
Simultaneously, the box length shrinks to \SI{24.76}{\AA}.
When calculating the new unwrapped coordinates, the wrapped x-coordinate for atom $a_1$ is moved by $\left\lfloor \frac{463.43-6.43}{24.76}+\frac{1}{2}\right\rfloor = \left\lfloor 18.96\right\rfloor =18$ PBC dimensions.
However, atom $a_2$ is moved by $\left\lfloor \frac{464.38-5.46}{24.76}+\frac{1}{2}\right\rfloor = \left\lfloor 19.03\right\rfloor =19$ PBC dimensions.
In this demonstration, the molecule is not near a box edge in wrapped coordinates, and so crossing events themselves are not problematic.
Instead, the error arises from mixing together coordinates expressed in different unit cell dimensions (\autoref{fig:schemata}).
More generally, in the limit when $x_i^u$ becomes large relative to $L_{i+1}$, small deviations in the box dimension can change the value for the floor function within Eq.~\ref{eq:heuristic} more or less at random.
Common barostats scale the atomic positions when the unit cell changes size\cite{Feller1995,Berendsen1984}, and this is not taken into account in Eq.~\ref{eq:heuristic}.
Thus, in rare circumstances the heuristic method miscounts the number of simulation boxes to move an atom within our trajectories (\autoref{fig:floor_error}A), in agreement with the findings by \citet{VonBulow2020}.
Since the unwrapped coordinates are cumulative with prior unwrapped steps, these errors propagate across time when unwrapping a simulation trajectory.
When the box size changes quickly, as is the case for our NAMD trajectories with a fluctuating barostat (\autoref{fig:si_fluctuations}), the floor function will more frequently split apart molecules (\autoref{fig:bonds}B).

Unlike the quantized heuristic errors, the bond length distribution for GROMACS trajectories handled with the displacement scheme is effectively continuous (\autoref{fig:bonds}A).
The displacement method does not miscount unit cell traversals.
Each unwrapping timestep effectively checks if a box jump event occurred and shifts the coordinates accordingly by at most one box vector.
However, the displacement scheme does not account for the coordinate scaling of the unwrapped position $x_i^u$ during the unwrapping process, and can under certain circumstances extend or shrink a bond by the difference in the unit cell dimensions.
When aggregating the results over many thousands of frames, the small bond length deviations add up to larger errors.
Interestingly, this is only a problem for the GROMACS trajectories.
GROMACS, unlike NAMD, will split molecules when wrapping trajectories across a periodic boundary.
Thus, the displacement vectors for GROMACS within a molecule can vary significantly between atoms from frame to frame.
NAMD, by contrast, wraps atoms within a molecule together by default, and so all atoms within a molecule will shift by the same amount, maintaining a consistent bond length.

To elucidate the displacement error, let us consider another concrete example from our simulation trajectory, again focusing on the behavior of two atoms.
After around 40 ps in the simulation, the x coordinates of two bound atoms $a_1$ and $a_2$ are $x_1=1.43\,\text{\normalfont\AA}$ and $x_2=0.92\,\text{\normalfont\AA}$, respectively.
The atoms did not move across the periodic box yet, and therefore the wrapped and unwrapped coordinates are identical.
The bond length between the atoms is around \SI{1}{\angstrom}, with an x-contribution of $\lvert 1.43-0.92 \rvert\,\text{\normalfont\AA} = 0.51\,\text{\normalfont\AA}$.
2 ps later, the two atoms moved backwards, resulting in the jump of $a_1$ across the periodic box.
The new wrapped coordinates are $x^w_1=24.54\,\text{\normalfont\AA}$ and $x_2=0.27\,\text{\normalfont\AA}$, with a box size of \SI{25.13}{\AA}, implying a bond length along the x-dimension of \SI{0.86}{\AA}.
Using the displacement unwrapping scheme from Eq.~\ref{eq:displacement} to unwrap the coordinates of $a_1$, the displacement between the current frame and the previous frame is added to the unwrapped coordinates from the previous frame, and the current box dimension is subtracted to account for the periodic jump that particle 1 undertook: $x^u_1=1.43\,\text{\normalfont\AA}+(24.54-1.43)\,\text{\normalfont\AA}-25.13\,\text{\normalfont\AA} = -0.59\,\text{\normalfont\AA}$.
At this point, the bond length along the x-dimension remains \SI{0.86}{\AA} for the unwrapped coordinates ($\lvert -0.59 - 0.27 \rvert\,\text{\normalfont\AA} = 0.86\,\text{\normalfont\AA}$).
However, after an additional 2 ps simulation time, the atoms moved again to the wrapped coordinates $x^w_1=24.47\,\text{\normalfont\AA}$ and $x_2=0.23\,\text{\normalfont\AA}$, and the box shrank to \SI{25.02}{\AA}.
Applying again the displacement unwrapping scheme to $a_1$ yields $x^u_1=-0.59\,\text{\normalfont\AA}+(24.47-24.54)\,\text{\normalfont\AA} = -0.66\,\text{\normalfont\AA}$.
Finally, we can compare the x contribution to the bond length again, yielding $\lvert -0.66 - 0.23 \rvert\,\text{\normalfont\AA} = 0.89\,\text{\normalfont\AA}$ for the unwrapped and $\lvert ( 24.47-25.02 ) - 0.23 \rvert\,\text{\normalfont\AA} = 0.78\,\text{\normalfont\AA}$, for the wrapped coordinates.
The barostat scaling the atomic coordinates when adjusting the box volume causes this discrepancy.
The error between the unwrapped and wrapped x coordinate difference, $\lvert 0.89 - 0.78 \rvert\,\text{\normalfont\AA} = 0.11\,\text{\normalfont\AA}$ is exactly the difference in the box dimension between the current and last frame $\lvert 25.02 - 25.13 \rvert\,\text{\normalfont\AA} = 0.11\,\text{\normalfont\AA}$.

\begin{figure}[h]
	\centering
	\includegraphics[width=0.45\linewidth]{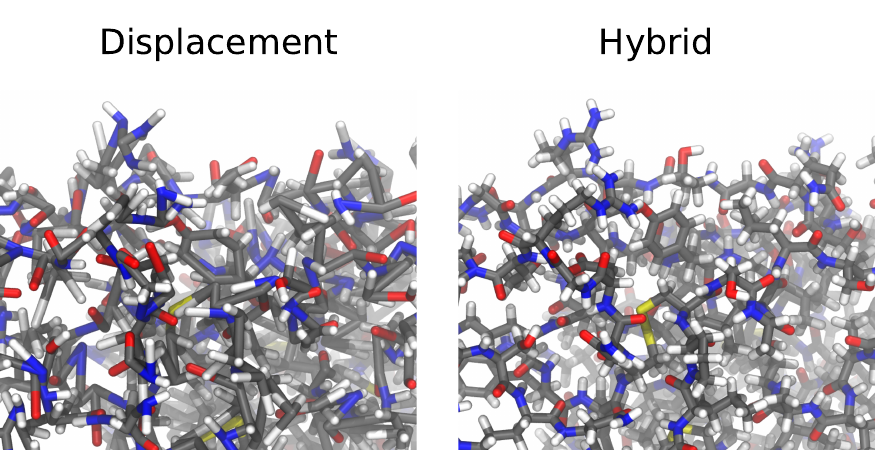}
	\caption{Lysozyme (pdb code 193L)\cite{Vaney1996} simulated for 1 $\mathrm{\mu s}$ in GROMACS and unwrapped with the (left) displacement and (right) hybrid method.
	The protein was represented by the AMBER ff99SB force field\cite{Lindorff-Larsen2010}. Atoms are colored by their element \textit{gray} carbon, \textit{red} oxygen, \textit{white} hydrogen, \textit{blue} nitrogen and \textit{yellow} sulfur.}
	\label{fig:vmd_error}
\end{figure}

Recognizing this error yields Eq.~\ref{eq:correction}, which was determined first empirically by noticing bond length changes for molecules near periodic boundaries caused by Eq.~\ref{eq:displacement}.
Eq.~\ref{eq:correction} handles changes to the periodic box dimensions explicitly, and as a result maintains the water geometry after unwrapping (\autoref{fig:bonds}).
While we only quantify the geometry distortions within a water molecule for the original displacement scheme, we would anticipate that intermolecular distances could also be distorted, particlarly for elements of the system that move quickly and cross many periodic boundaries.
For GROMACS trajectories, the effect is more severe, as even the unwrapping of equilibrium protein simulations with the displacement scheme visually generates distorted and unphysical conformations (\autoref{fig:vmd_error}).
While pre-joining connected atoms may correct intramolecular geometries, intermolecular distances may still change, yielding incorrect relative geometries between molecules.

A related consideration here is the estimate for the error terms.
As demonstrated in the derivation, the heuristic scheme fundamentally assumes that $| \frac{d}{L_{i+1}} |<\frac{1}{2}$.
From \autoref{fig:floor_error}A, we see that this is not always true, particularly when the particle travels far from the origin (\autoref{fig:floor_error}B).
However, following the assumption we make to simplify \autoref{eq:t_displacement_2} ($|  m\left(\frac{ L_{i}-L_{i+1}}{L_{i+1}}\right) - \frac{d}{L_{i+1}} |<\frac{1}{2}$), we see that $m\left(\frac{ L_{i}-L_{i+1}}{L_{i+1}}\right) $ is correlated to $\frac{d}{L_{i+1}}$ (\autoref{fig:floor_error}C).
Thus, for cases where the key assumption for the heuristic scheme fails, the requirement for the hybrid scheme is still valid (\autoref{fig:floor_error}D).

\subsection{Scheme Reversibility}

\newcommand{\posTable}{Time	&	\multicolumn{3}{c}{RMSD in \AA}	\\
	&	Heuristic	&	Disp	&	Hybrid	\\
0 fs	&	0	&	0	&	0	\\
\rowcolor[HTML]{CCCCCC}
0.2 ps	&	0	&	0	     &	$6 \cdot 10^{-8}$	\\
1 ps	&	0	&	0.036	 &	$8 \cdot 10^{-8}$	\\
\rowcolor[HTML]{CCCCCC}
10 ps	&	0	&	0.042	 &	$2 \cdot 10^{-7}$	\\
0.1 ns	&	0	&	0.145	 &	$10^{-6}$	\\
\rowcolor[HTML]{CCCCCC}
1 ns	&	0	&	0.57	 &	$3 \cdot 10^{-6}$	\\
10 ns	&	0	&	1.9	    &	$8 \cdot 10^{-6}$	\\
\rowcolor[HTML]{CCCCCC}
0.1 μs	&	4.95	&	5.8	&	$10^{-5}$	\\
1 μs	&	184	&	18	    &	$2 \cdot 10^{-5}$ \\
}
\begin{table}[!ht]
	\begin{tabular}[width=\textwidth]{lccc}
		Time	&	\multicolumn{3}{c}{RMSD in \AA}	\\
	&	Heuristic	&	Disp	&	Hybrid	\\
0 fs	&	0	&	0	&	0	\\
\rowcolor[HTML]{CCCCCC}
0.2 ps	&	0	&	0	     &	$6 \cdot 10^{-8}$	\\
1 ps	&	0	&	0.036	 &	$8 \cdot 10^{-8}$	\\
\rowcolor[HTML]{CCCCCC}
10 ps	&	0	&	0.042	 &	$2 \cdot 10^{-7}$	\\
0.1 ns	&	0	&	0.145	 &	$10^{-6}$	\\
\rowcolor[HTML]{CCCCCC}
1 ns	&	0	&	0.57	 &	$3 \cdot 10^{-6}$	\\
10 ns	&	0	&	1.9	    &	$8 \cdot 10^{-6}$	\\
\rowcolor[HTML]{CCCCCC}
0.1 μs	&	4.95	&	5.8	&	$10^{-5}$	\\
1 μs	&	184	&	18	    &	$2 \cdot 10^{-5}$ \\

	\end{tabular}
	\caption{\textmd{Average differences in atomic positions between unwrapping methods and reference after certain simulation times. 
		The reference positions are obtained from a NAMD trajectory without wrapping.
		To get coordinates for the different unwrapping methods, the reference trajectory is wrapped and subsequently unwrapped by the respective method.}}
	\label{tab:positions}
\end{table}

The application of wrapping and unwrapping should be a reversible operation that might be applied repeatedly, such as when a multi-domain protein is made whole via unwrapping prior to rewrapping the trajectory for analysis.
To test the reversibility for the schemes introduced by Eqs.~\ref{eq:heuristic}, \ref{eq:displacement} and \ref{eq:correction}, the NAMD unwrapped NPT reference trajectory was wrapped followed by unwrapping with the respective method.
Afterwards, the RMSD between the original and newly unwrapped coordinates were compared (\autoref{tab:positions}).
Both the heuristic and displacement scheme did not preserve atomic positions, with significant RMSDs for the displacement method occurring after 0.1\,ns, and the heuristic method failing after 100\,ns when the atoms travel far from the origin.
By contrast, the hybrid method preserves atomic positions to within the limit of single precision floating point arithemtic, with the increase in RMSD corresponding to the larger and larger atomic positions that reduce accuracy.

The key advance that enables the success for Eq.~\ref{eq:correction} is the $m (L_{i+1}-L_{i})$ term that takes into account the scaling of atomic positions $x_i^u$ after a change in the box volume.
The traversed boxes count is multiplied with the box size change between the current $L_{i+1}$ and last $L_{i}$ iteration, and subtracted from the atom positions as defined in Eq.~\ref{eq:correction}.
The additional term corrects the diffusion overestimate for the heuristic scheme (Eq.~\ref{eq:heuristic}) observed in \autoref{fig:msd}.
As seen in \autoref{fig:bonds}, the bond lengths when applying \autoref{eq:correction} are uniform, and their \SI{0.1}{\nano\meter} length conforms to our expectation for water structure.

Table~\ref{tab:positions} also suggests that the correct choice for unwrapping schemes depends on the context.
The heuristic scheme's shortcomings only become readily apparent in a long simulation where the box size varies.
Unwrapping NVT simulations would be expected to yield correct results under any scheme, so long as trajectory frames are taken frequently enough so that PBC crossing events are not missed.
The lower memory footprint for the heuristic scheme may be beneficial in some instances.
However, when unwrapping a trajectory with changing box dimensions, as is typically the case for simulations in a constant pressure ensemble, the heuristic scheme will eventually fail to accurately unwrap cordinates.
In these instances, the hybrid scheme from Eq.~\ref{eq:correction} is essential to capturing the correct molecular geometries and diffusive behavior.
To facilitate wide adoption for the new scheme, we have implemented Eq.~\ref{eq:correction} into the VMD\cite{Humphrey1996} packages \href{https://github.com/kulkem/pbctools.git}{pbctools}, fastpbc and \href{https://github.com/kulkem/qwrap.git}{qwrap}\cite{Henin2020}.
\section{Conclusion}

In this study, we have demonstrated that the current implementations for the heuristic and displacement unwrapping methods lead to inaccurate unwrapped atomic positions through three different criteria; measured diffusion coefficients, bond lengths, and reversibility for wrapping and unwrapping procedures.
The heuristic case (Eq.~\ref{eq:heuristic}) struggles if the simulated time scale leads to atomic positions that would be much larger than the periodic box size, leading to uncertainties for how many periodic boxes a molecule needs to be moved when the PBC dimensions change.
The displacement method (Eq.~\ref{eq:displacement}) leads to changing internal geometries as the box dimensions change, and also perturbs intermolecular distances after rewrapping the trajectories.
We therefore proposed a hybrid scheme Eq.~\ref{eq:correction} that faithfully reproduces molecular geometries over time as periodic cell dimensions change, and recommend its adoption when analyzing future simulations.

However, past conclusions from these simulations are unlikely to materially change, irrespective for the wrapping algorithm.
Prior unwrapping schemes fail most obviously for pathological systems where a  small fluctuating unit cell is simulated for an extended period of time.
Only in a handful of studies likely meet these criteria, as diffusion coefficient estimates converge quickly enough that extended simulations are not required.
Likewise, the trend towards larger simulation systems increases the timescales over which the heuristic approximation from \autoref{eq:heuristic} is valid.
However, with the increasing availability for graphical processing units\cite{Phillips2020,Pall2020} or special purpose machines such as Anton 3\cite{Shaw2021} to deliver longer and longer trajectories, we foresee a future where box dimension changes must explicitly be accounted during trajectory analysis for accurate results.

\section{Acknowledgements}
The authors thank Michigan State University (MSU) for providing startup funds. 
This work used the Extreme Science and Engineering Discovery Environment (XSEDE), which is supported by National Science Foundation grant number ACI-1548562\cite{Towns2014} (project number TG-BIO210040).
This work was supported in part through computational resources and services provided by the Institute for Cyber-Enabled Research at Michigan State University.
\section{Disclosure}
The authors report no conflicts of interest in this work.

\suppinfo
The Supporting Information includes 33 figures referenced in the text.
Figures S1-S31 compute the mean square displacement for water molecules within individual simulation systems, and use linear fits to calculate diffusion coefficients.
\autoref{fig:si_fluctuations} measures the fluctuations for the unit cell dimensions between our NAMD and GROMACS trajectories.
\autoref{fig:si_disphist} measures the displacement histograms within our unwrapped NAMD trajectory.

\bibliography{misc/library.bib}

\end{document}


\subsection{This PDF file includes}
\begin{itemize}
	\item Figures S1 to S33
\end{itemize}

\subsection{Other supplementary information}
\begin{itemize}
	\item Dataset for the simulation on \href{https://www.researchgate.net/publication/356424261_Data_set_for_Investigating_Reversibility_in_Unwrapping_Schemes_for_Molecular_Dynamics_Simulations}{ResearchGate}. Contains simulation input, topology and coordinate files to rerun the simulations, as well as python and tcl scripts to do the analysis and generate the images in this publication.
	\item \texttt{pbctools} implementation is on \href{https://github.com/kulkem/pbctools.git}{github}.
	\item \texttt{qwrap} implementation is also on \href{https://github.com/kulkem/qwrap.git}{github}.
	\item \texttt{fastpbc} implementation, which has already been integrated into the \href{http://www.ks.uiuc.edu/Research/vmd/doxygen/cvsget.html}{VMD source tree}.
\end{itemize}

\begin{figure*}[ht]
	\centering
	\includegraphics[width=\linewidth]{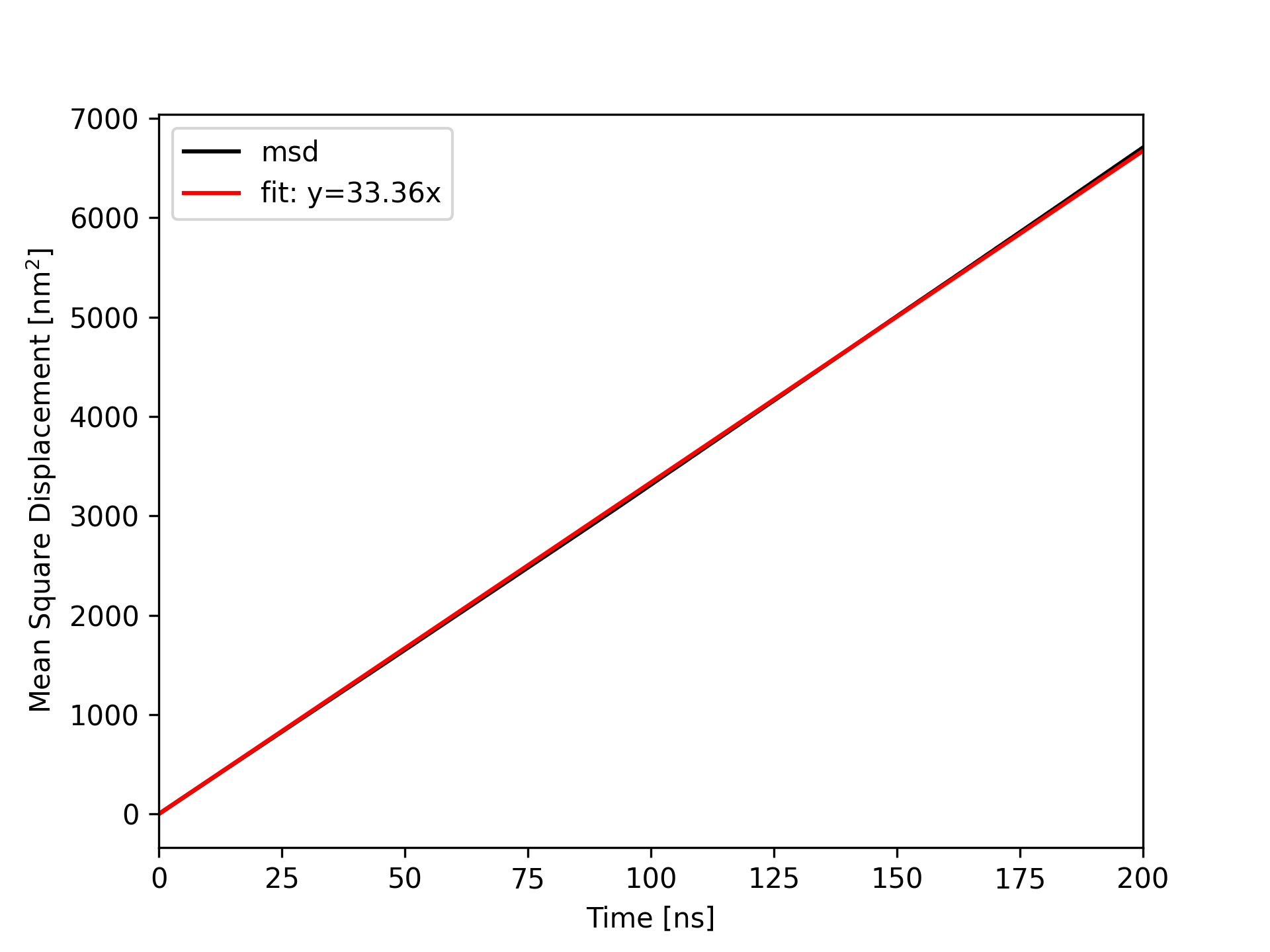}
   \caption{Mean square displacements for the NPT simulation in GROMACS with the displacement method and a trajectory frequency of \SI{2}{\pico\second}.}
\end{figure*}
\begin{figure*}[ht]
	\centering
	\includegraphics[width=\linewidth]{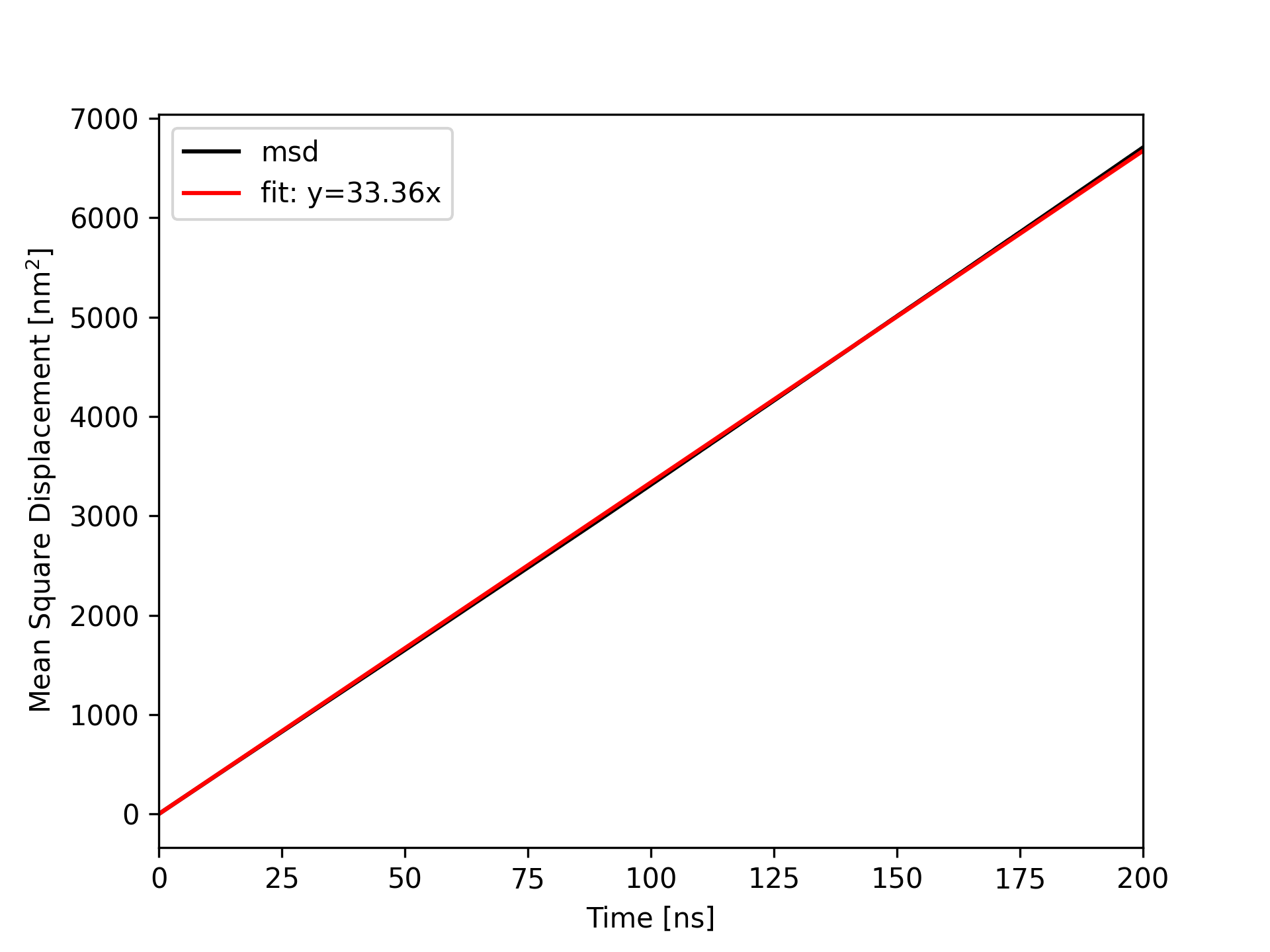}
   \caption{Mean square displacements for the NPT simulation in GROMACS with the displacement method and a trajectory frequency of \SI{1}{\pico\second}.}
\end{figure*}
\begin{figure*}[ht]
	\centering
	\includegraphics[width=\linewidth]{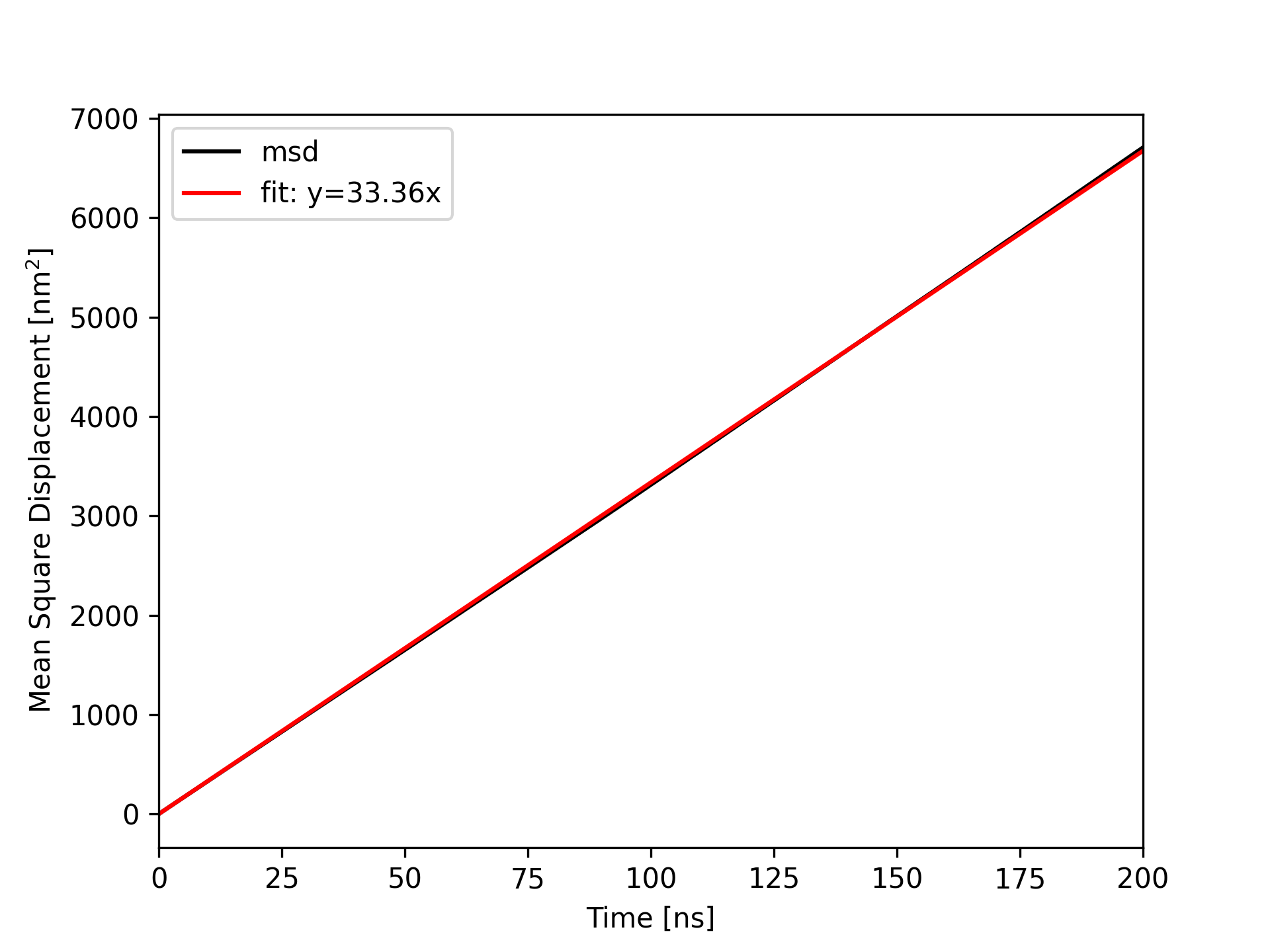}
   \caption{Mean square displacements for the NPT simulation in GROMACS with the displacement method and a trajectory frequency of \SI{0.5}{\pico\second}.}
\end{figure*}
\begin{figure*}[ht]
	\centering
	\includegraphics[width=\linewidth]{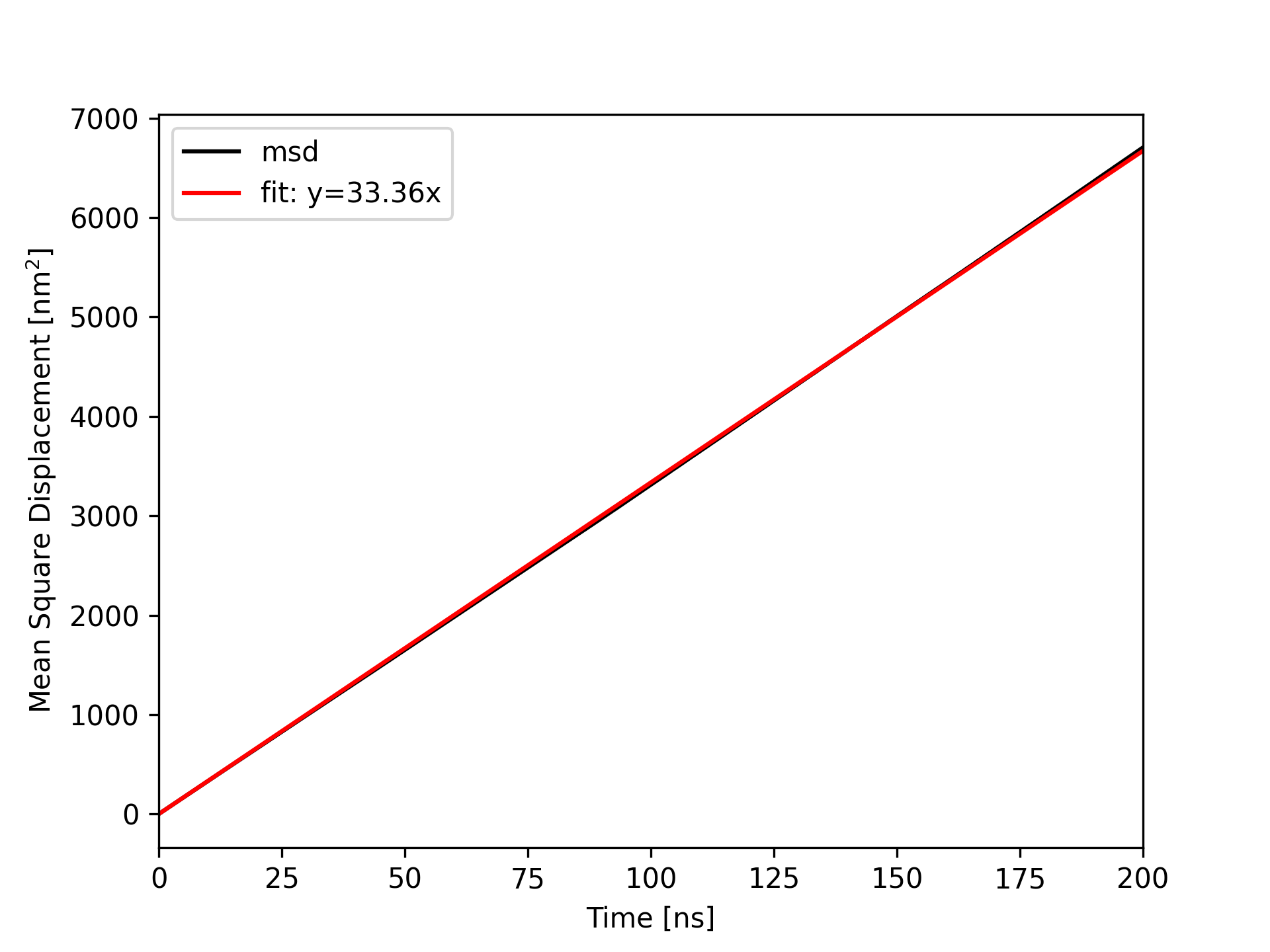}
   \caption{Mean square displacements for the NPT simulation in GROMACS with the displacement method and a trajectory frequency of \SI{0.2}{\pico\second}.}
\end{figure*}
\begin{figure*}[ht]
	\centering
	\includegraphics[width=\linewidth]{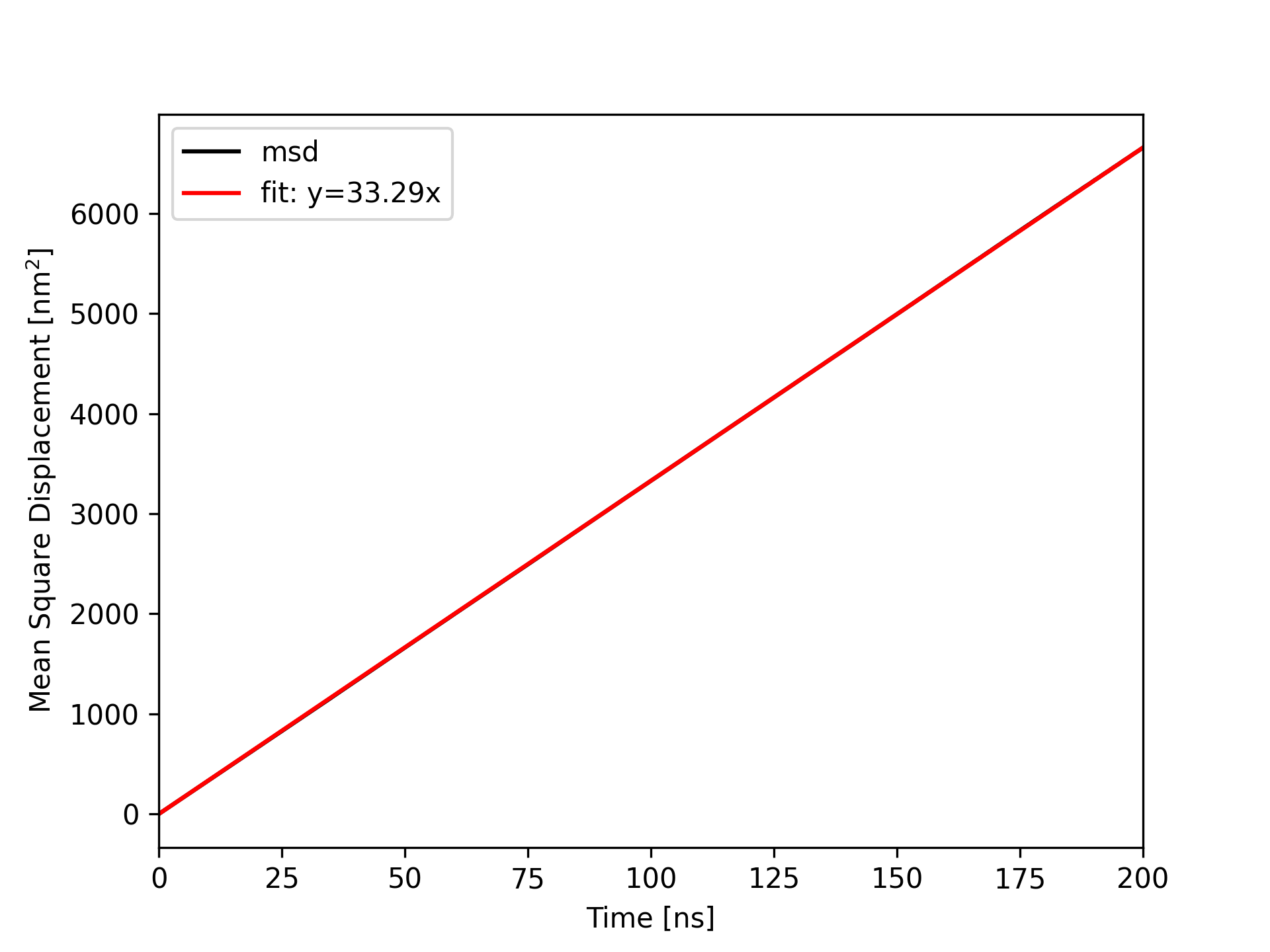}
   \caption{Mean square displacements for the NVT simulation in GROMACS with the displacement method and a trajectory frequency of \SI{0.2}{\pico\second}.}
\end{figure*}
\begin{figure*}[ht]
	\centering
	\includegraphics[width=\linewidth]{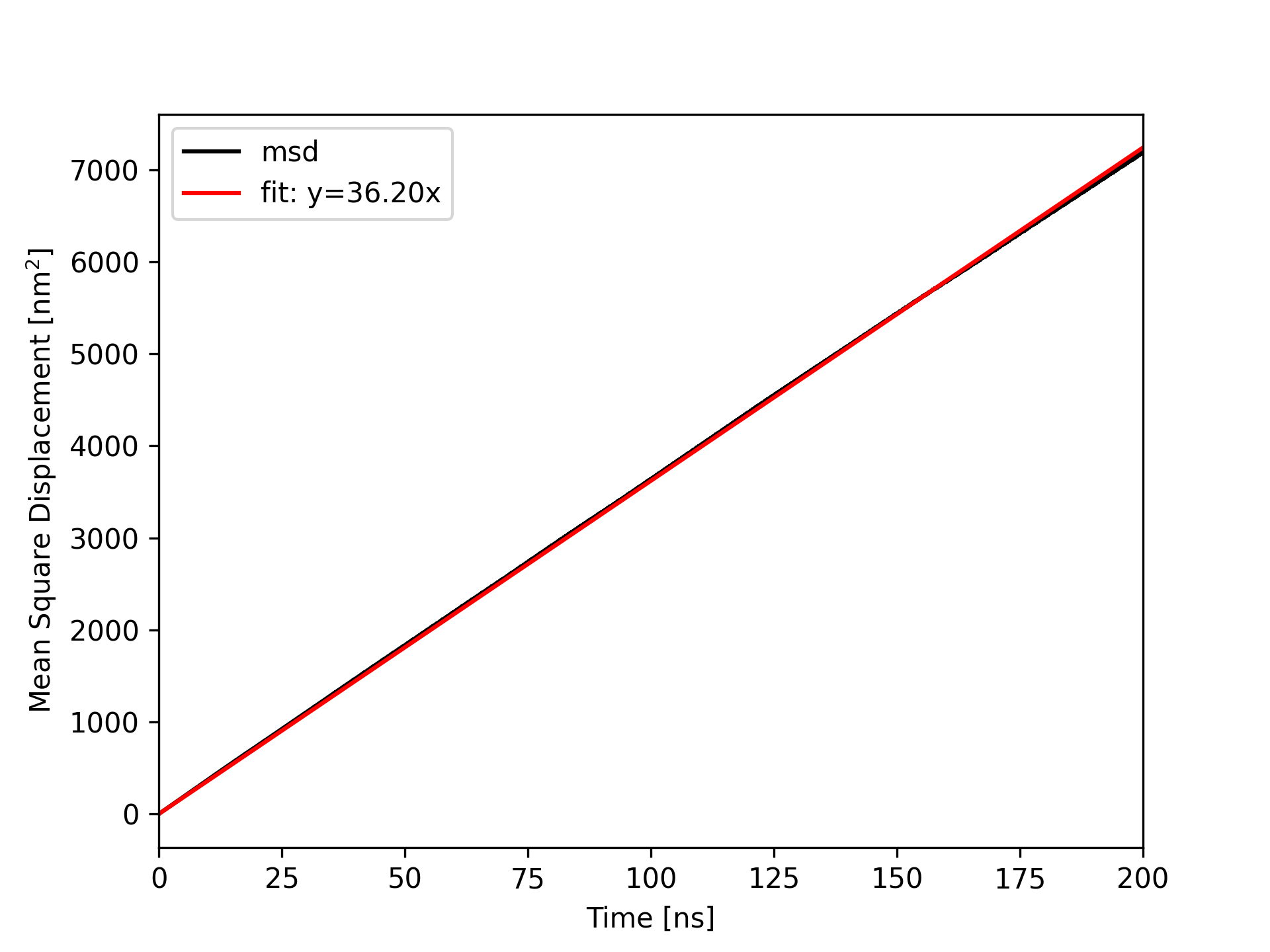}
   \caption{Mean square displacements for the NPT simulation in GROMACS with the heuristic method and a trajectory frequency of \SI{2}{\pico\second}.}
\end{figure*}
\begin{figure*}[ht]
	\centering
	\includegraphics[width=\linewidth]{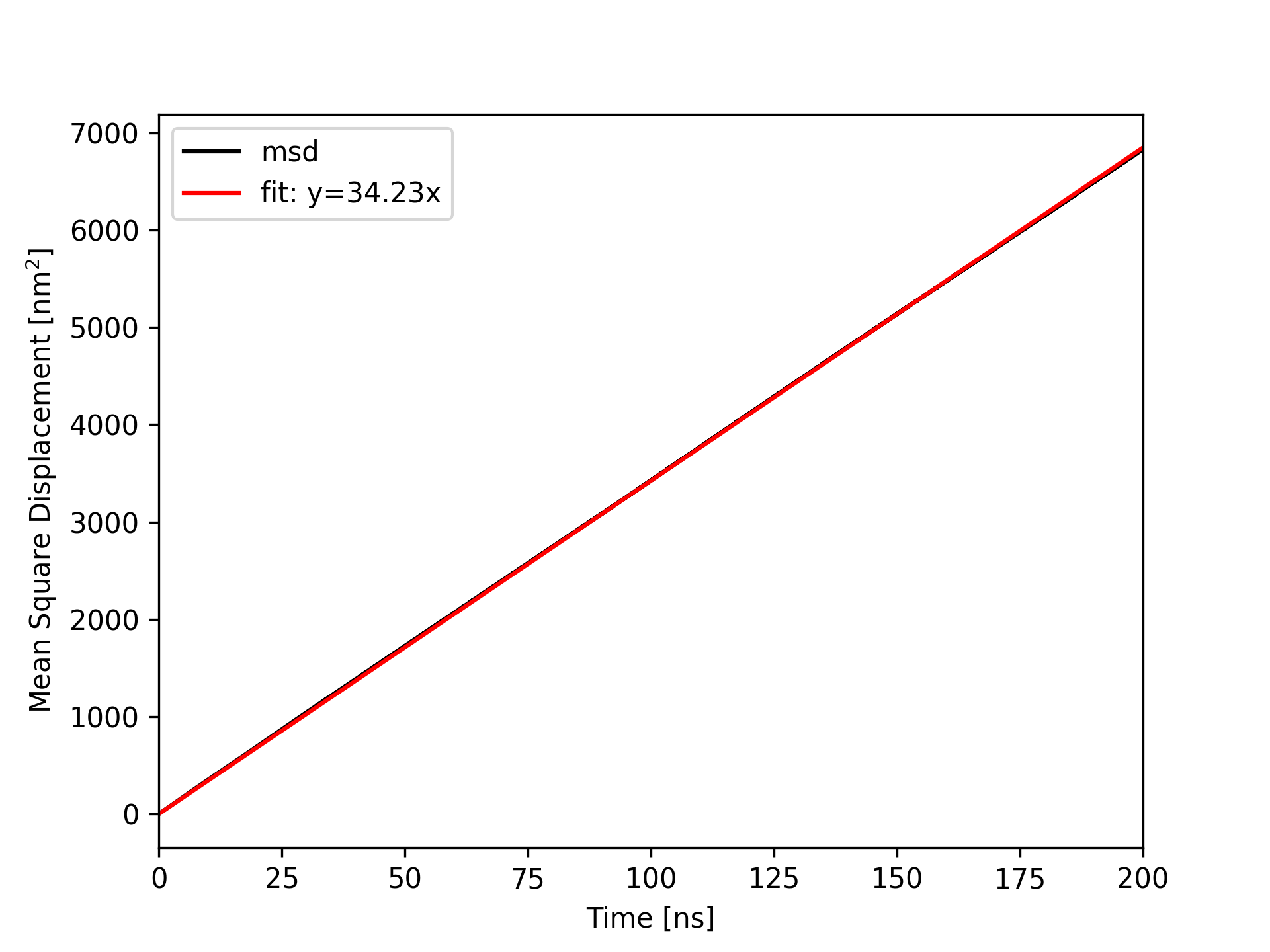}
   \caption{Mean square displacements for the NPT simulation in GROMACS with the heuristic method and a trajectory frequency of \SI{1}{\pico\second}.}
\end{figure*}
\begin{figure*}[ht]
	\centering
	\includegraphics[width=\linewidth]{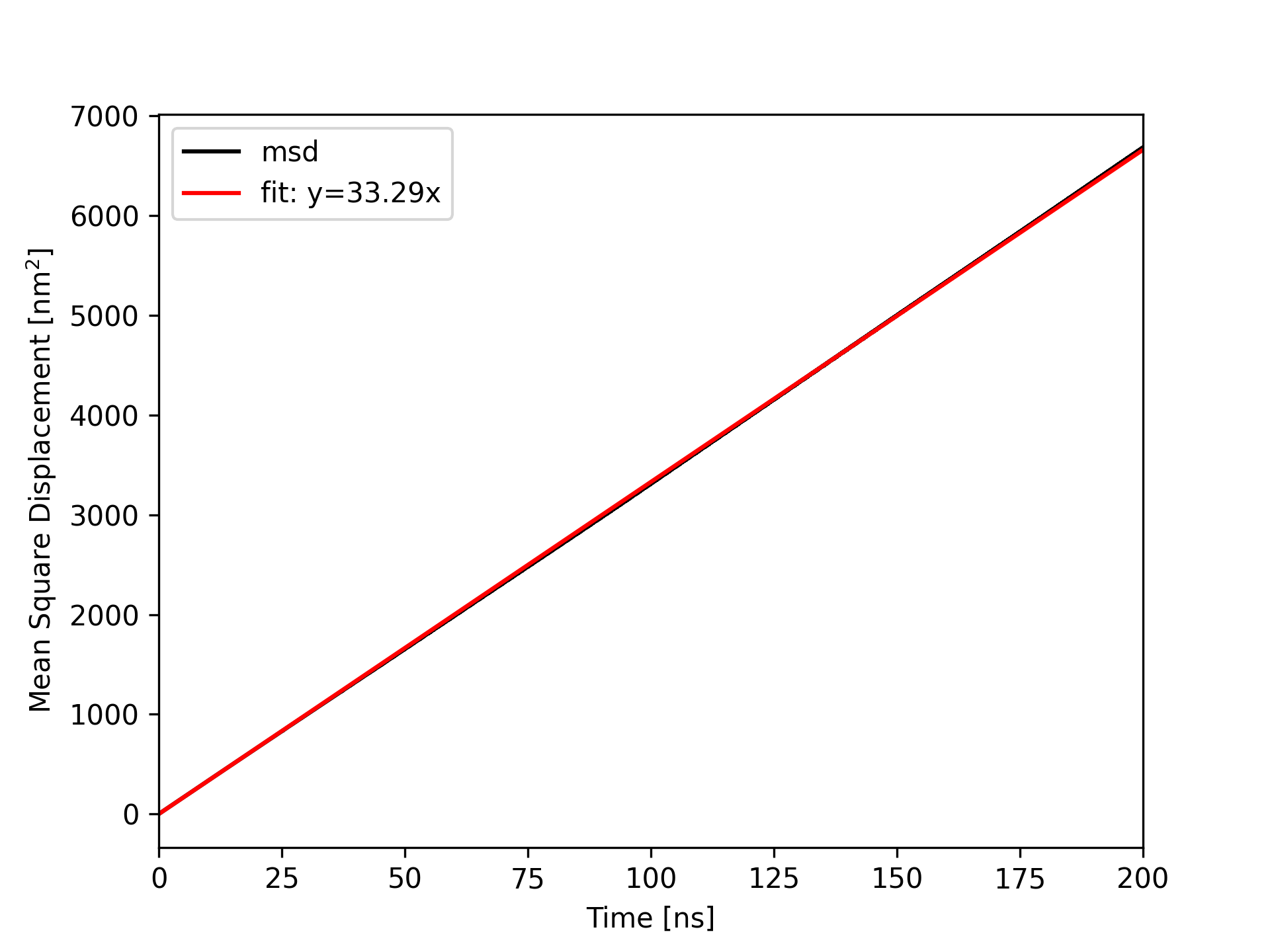}
   \caption{Mean square displacements for the NPT simulation in GROMACS with the heuristic method and a trajectory frequency of \SI{0.5}{\pico\second}.}
\end{figure*}
\begin{figure*}[ht]
	\centering
	\includegraphics[width=\linewidth]{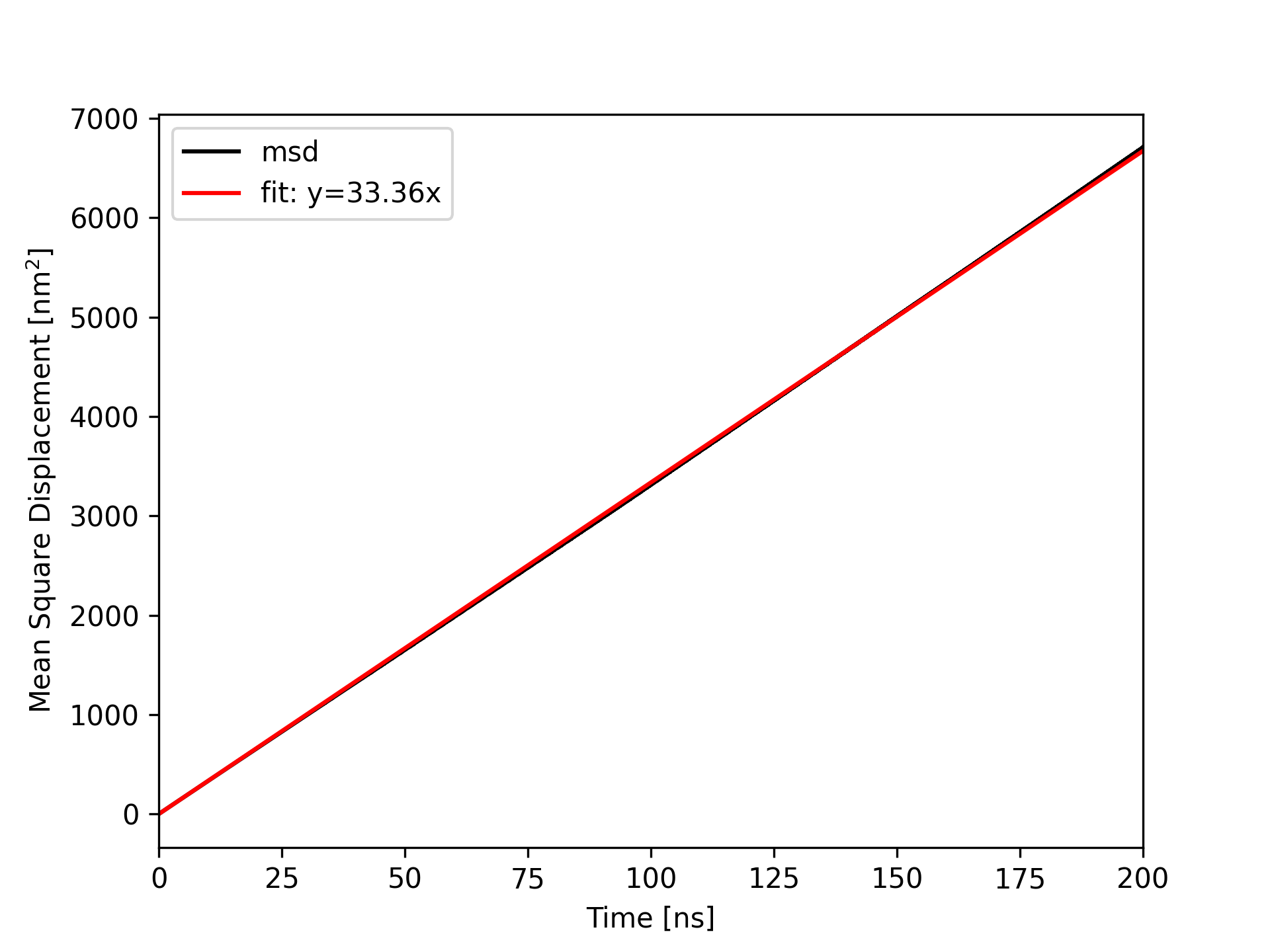}
   \caption{Mean square displacements for the NPT simulation in GROMACS with the heuristic method and a trajectory frequency of \SI{0.2}{\pico\second}.}
\end{figure*}
\begin{figure*}[ht]
	\centering
	\includegraphics[width=\linewidth]{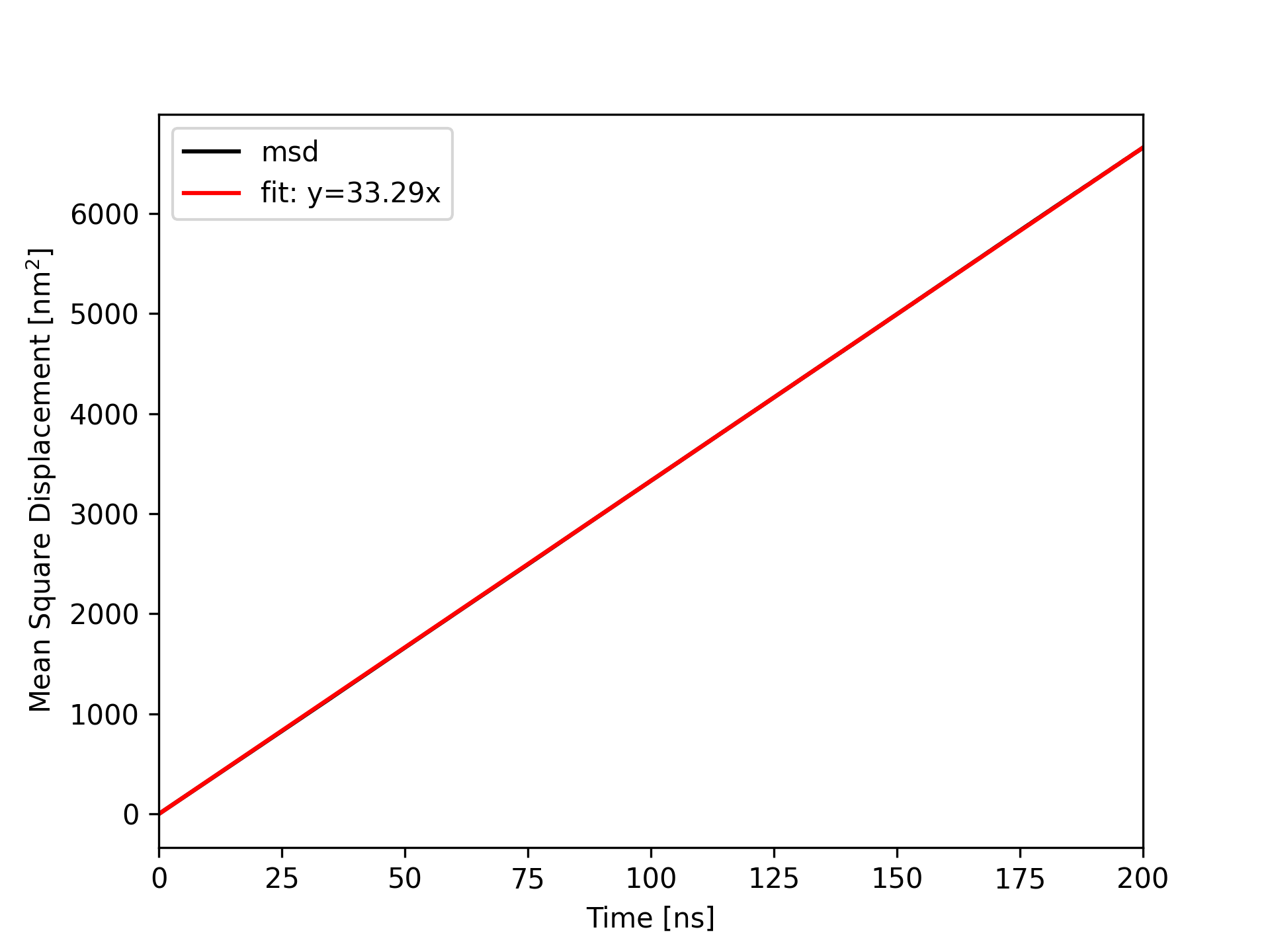}
   \caption{Mean square displacements for the NVT simulation in GROMACS with the heuristic method and a trajectory frequency of \SI{0.2}{\pico\second}.}
\end{figure*}
\begin{figure*}[ht]
	\centering
	\includegraphics[width=\linewidth]{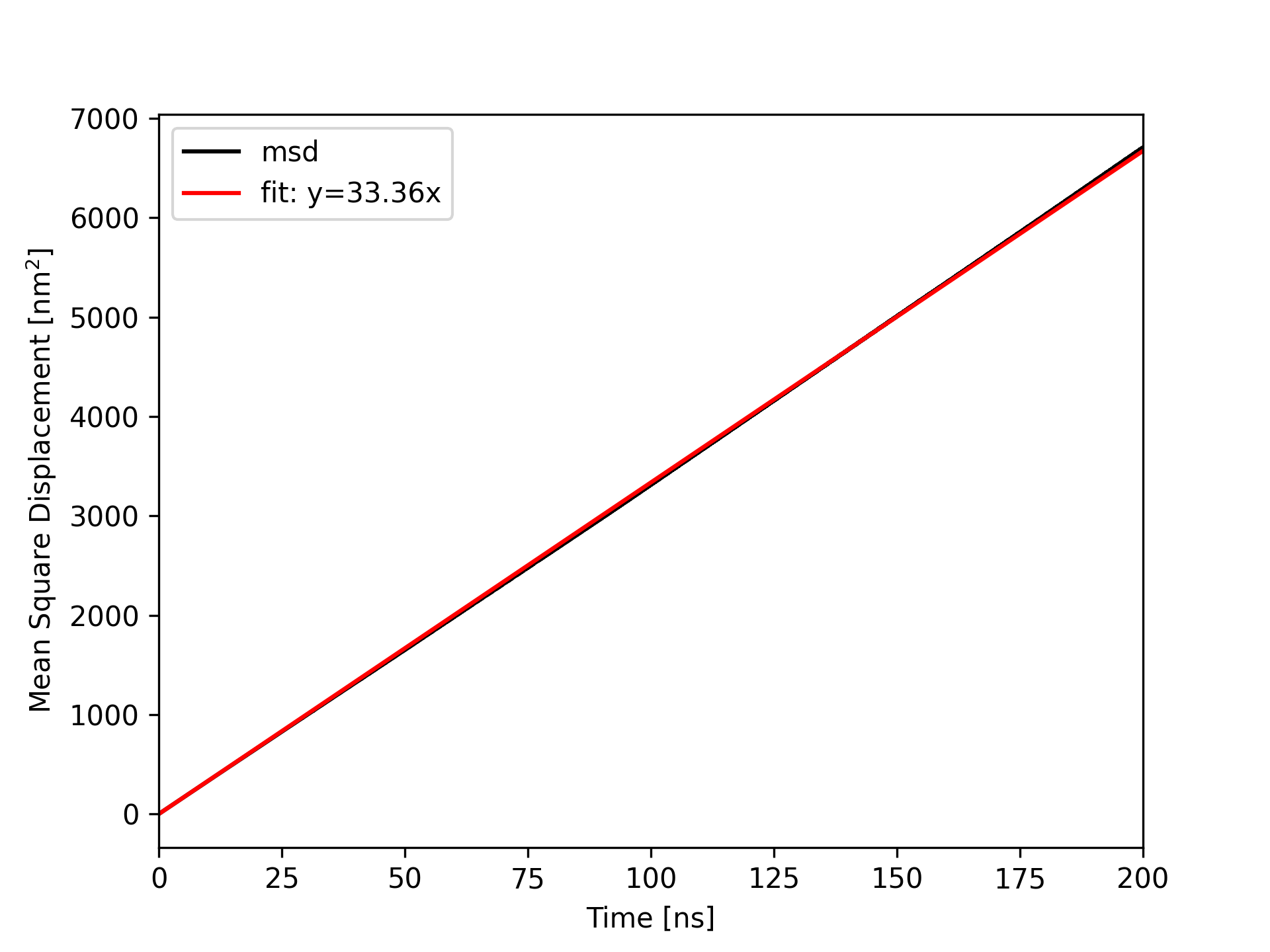}
   \caption{Mean square displacements for the NPT simulation in GROMACS with the hybrid method and a trajectory frequency of \SI{2}{\pico\second}.}
\end{figure*}
\begin{figure*}[ht]
	\centering
	\includegraphics[width=\linewidth]{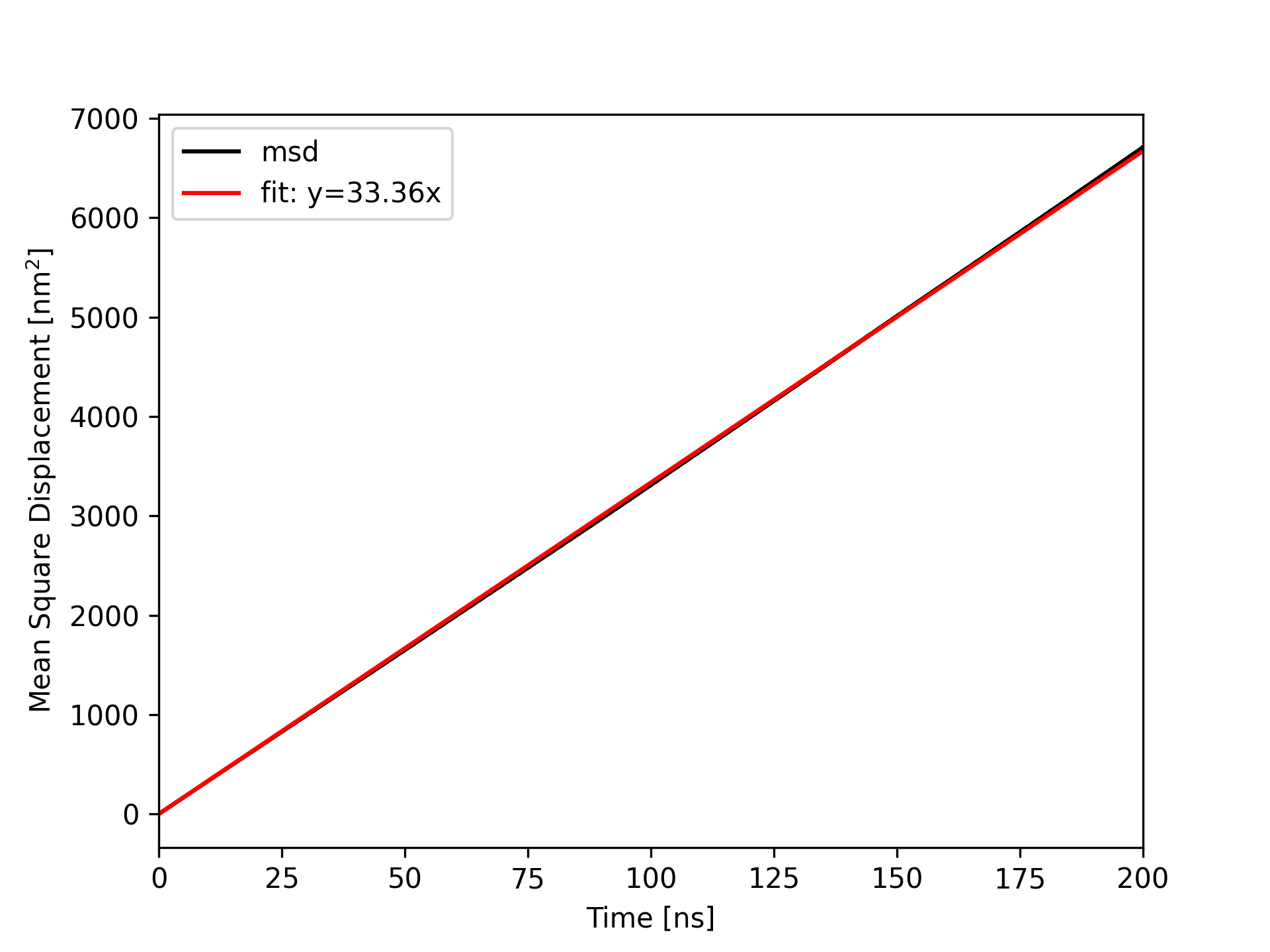}
   \caption{Mean square displacements for the NPT simulation in GROMACS with the hybrid method and a trajectory frequency of \SI{1}{\pico\second}.}
\end{figure*}
\begin{figure*}[ht]
	\centering
	\includegraphics[width=\linewidth]{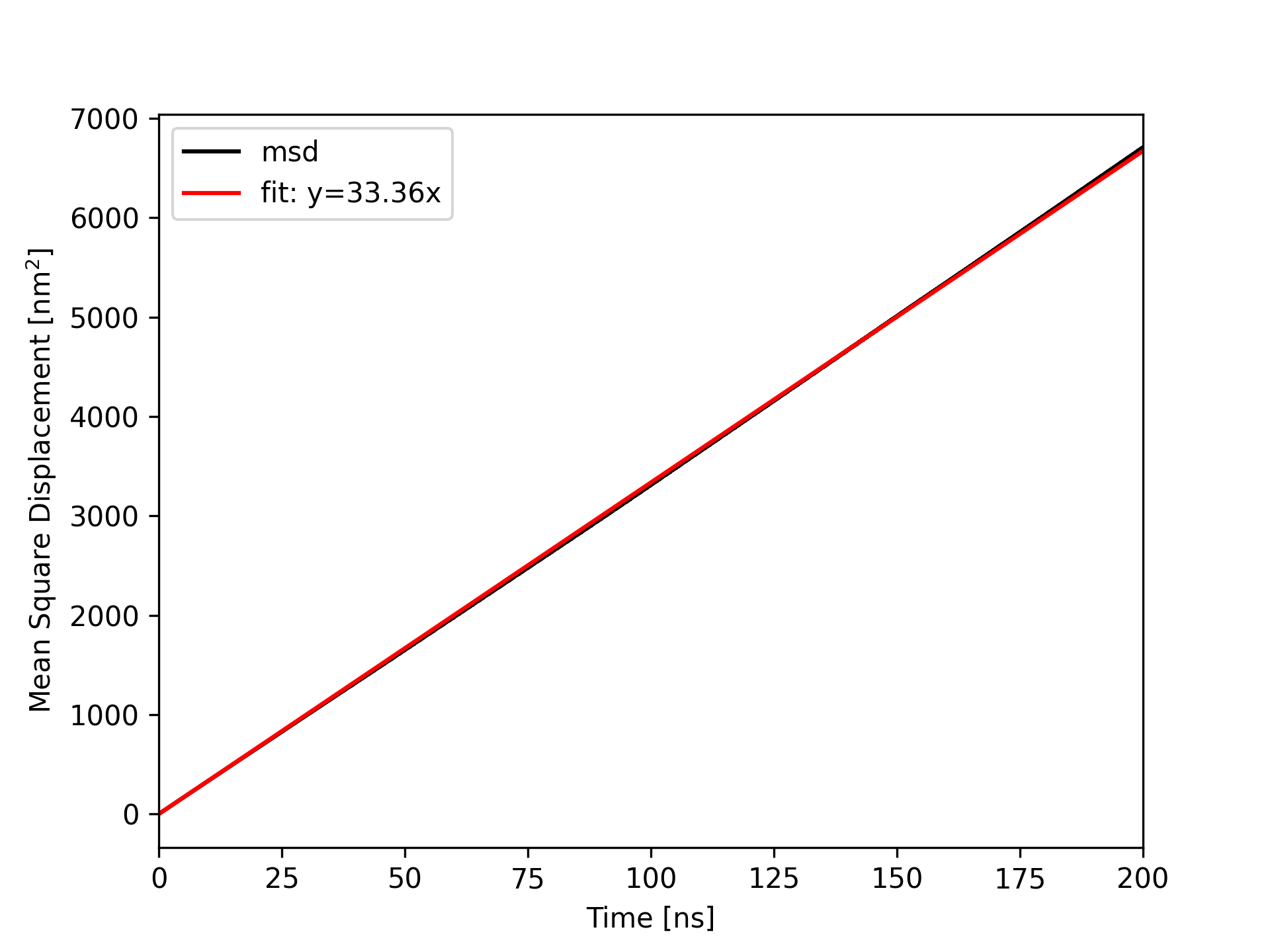}
   \caption{Mean square displacements for the NPT simulation in GROMACS with the hybrid method and a trajectory frequency of \SI{0.5}{\pico\second}.}
\end{figure*}
\begin{figure*}[ht]
	\centering
	\includegraphics[width=\linewidth]{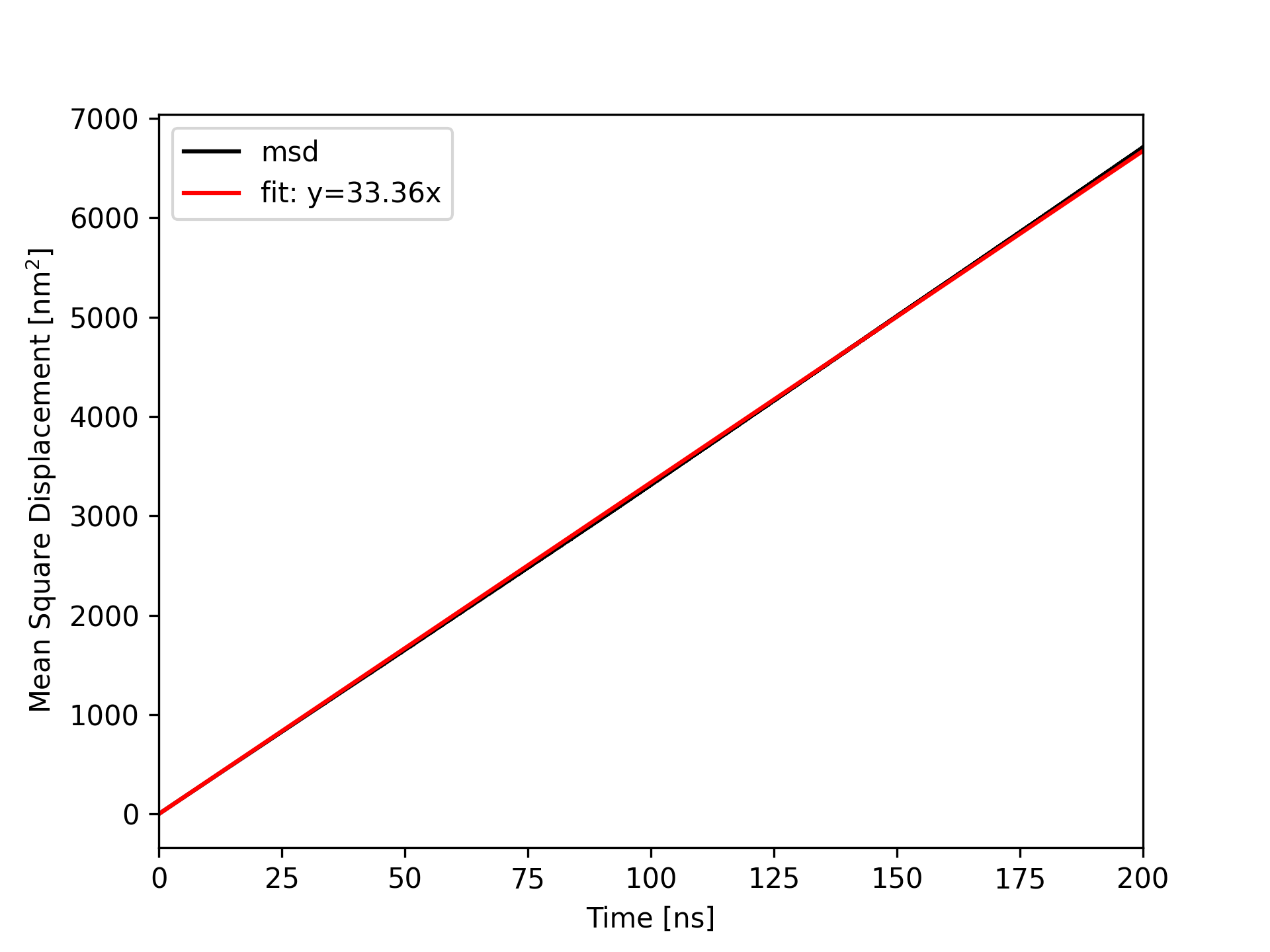}
   \caption{Mean square displacements for the NPT simulation in GROMACS with the hybrid method and a trajectory frequency of \SI{0.2}{\pico\second}.}
\end{figure*}
\begin{figure*}[ht]
	\centering
	\includegraphics[width=\linewidth]{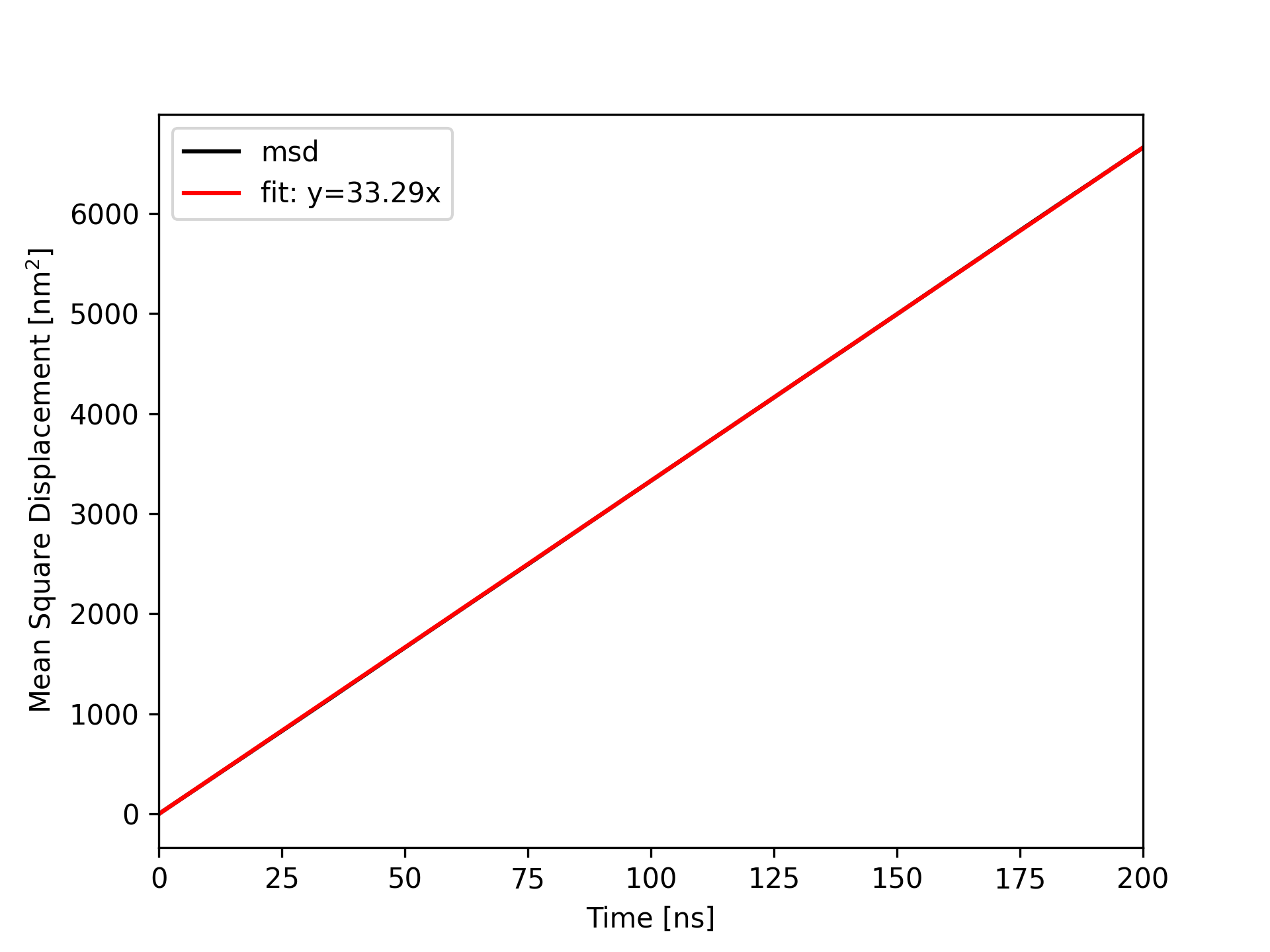}
   \caption{Mean square displacements for the NVT simulation in GROMACS with the hybrid method and a trajectory frequency of \SI{0.2}{\pico\second}.}
\end{figure*}
\begin{figure*}[ht]
	\centering
	\includegraphics[width=\linewidth]{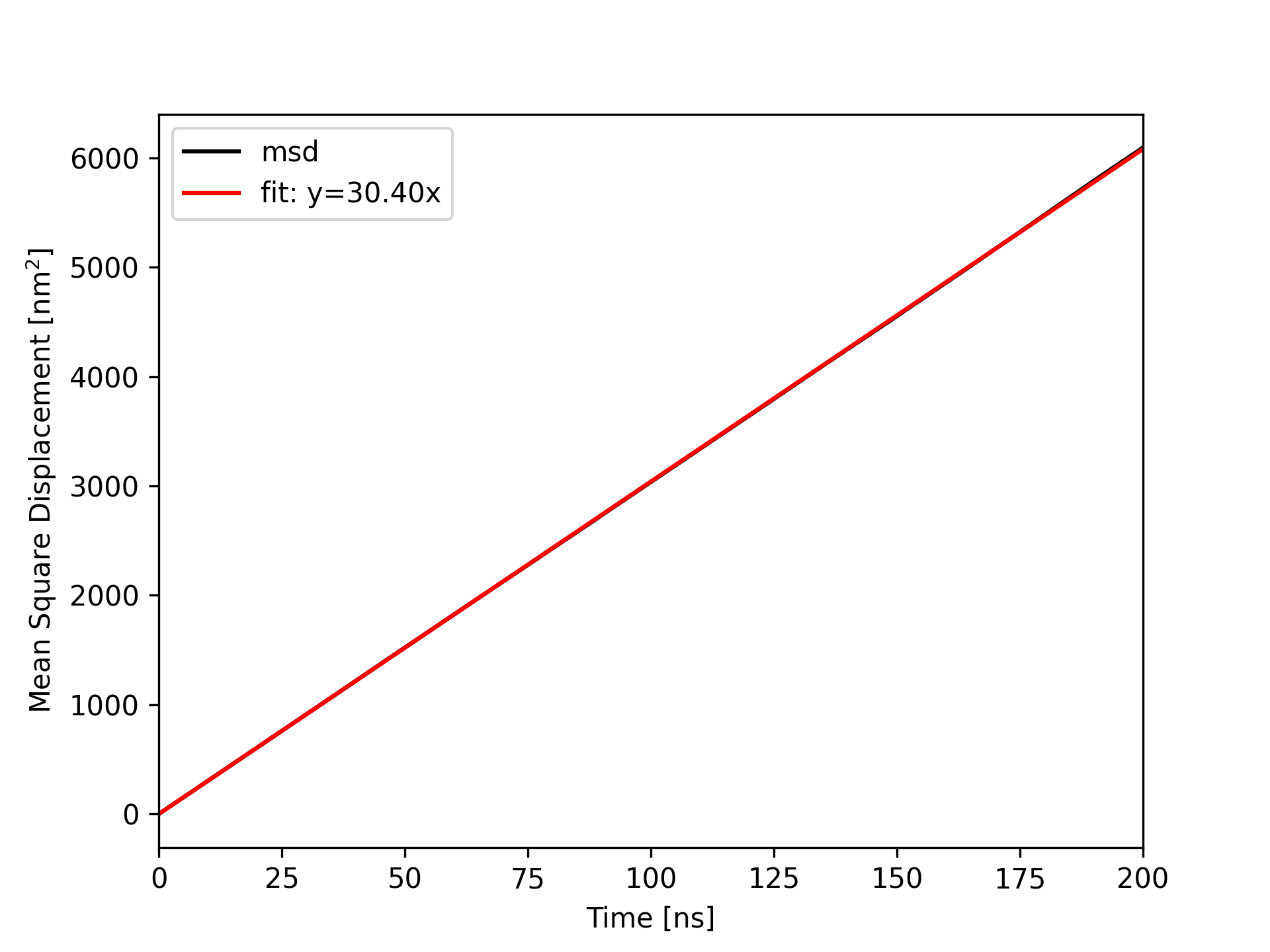}
   \caption{Mean square displacements for the NPT simulation in NAMD with the displacement method and a trajectory frequency of \SI{2}{\pico\second}.}
\end{figure*}
\begin{figure*}[ht]
	\centering
	\includegraphics[width=\linewidth]{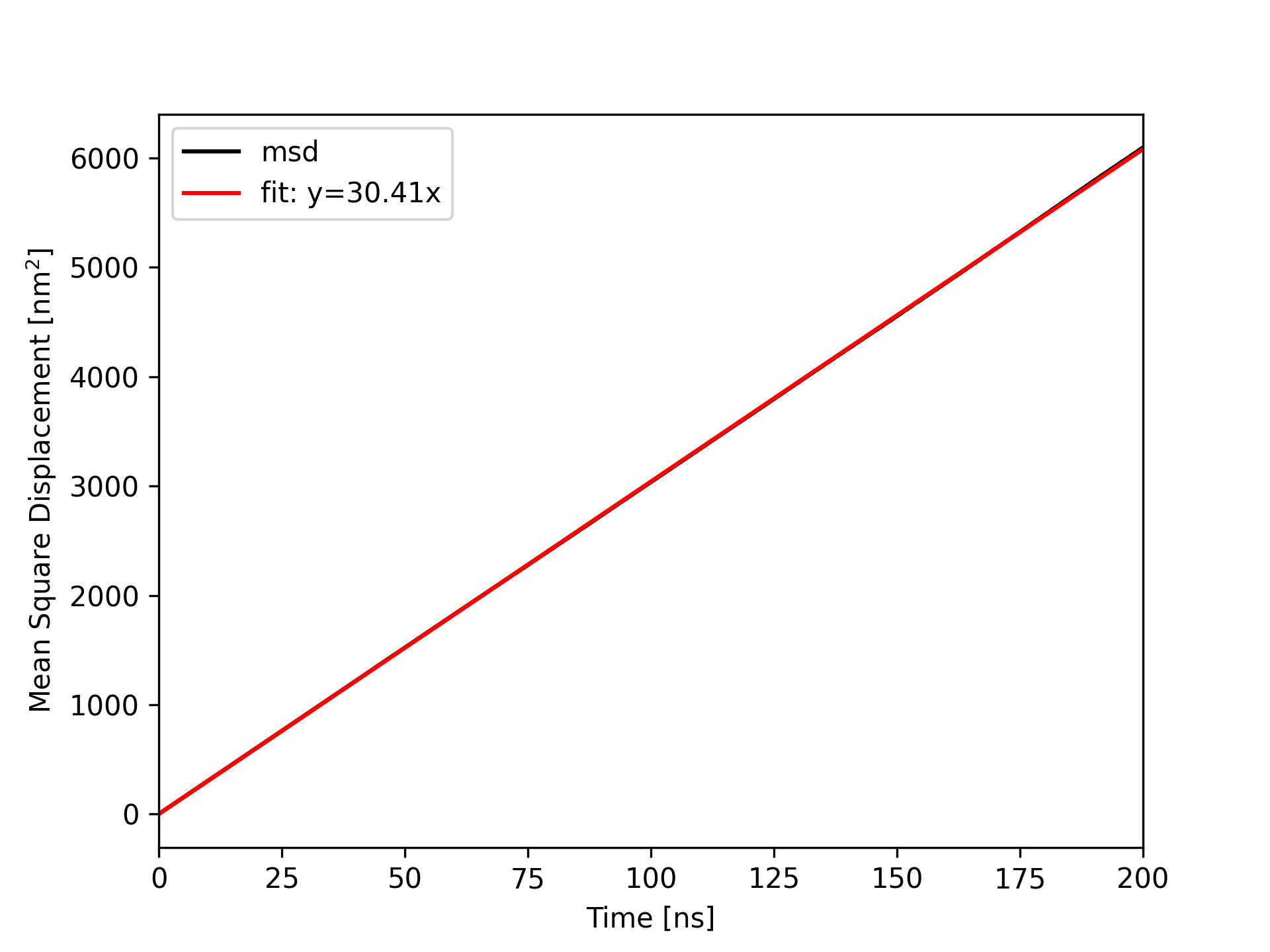}
   \caption{Mean square displacements for the NPT simulation in NAMD with the displacement method and a trajectory frequency of \SI{1}{\pico\second}.}
\end{figure*}
\begin{figure*}[ht]
	\centering
	\includegraphics[width=\linewidth]{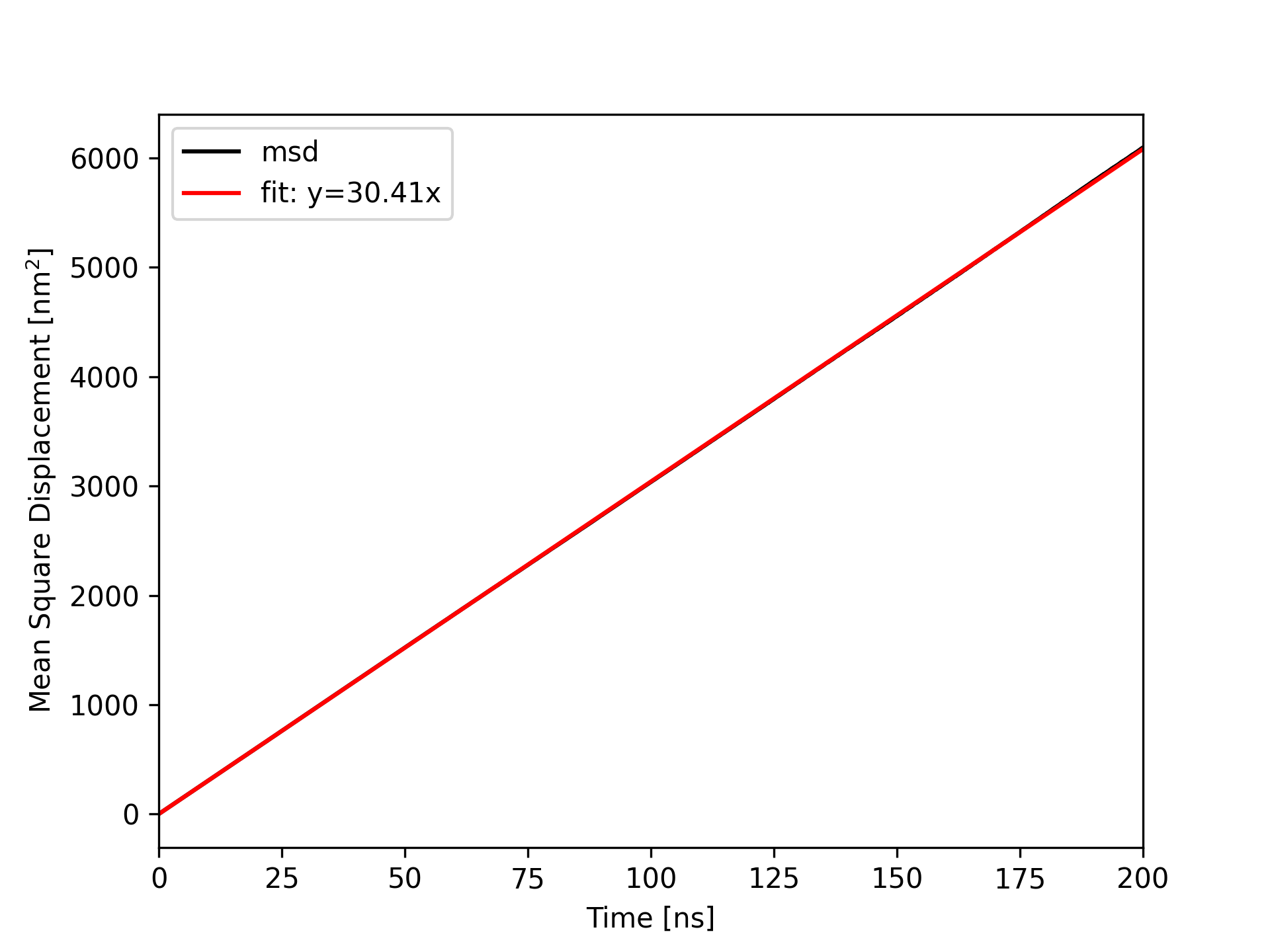}
   \caption{Mean square displacements for the NPT simulation in NAMD with the displacement method and a trajectory frequency of \SI{0.5}{\pico\second}.}
\end{figure*}
\begin{figure*}[ht]
	\centering
	\includegraphics[width=\linewidth]{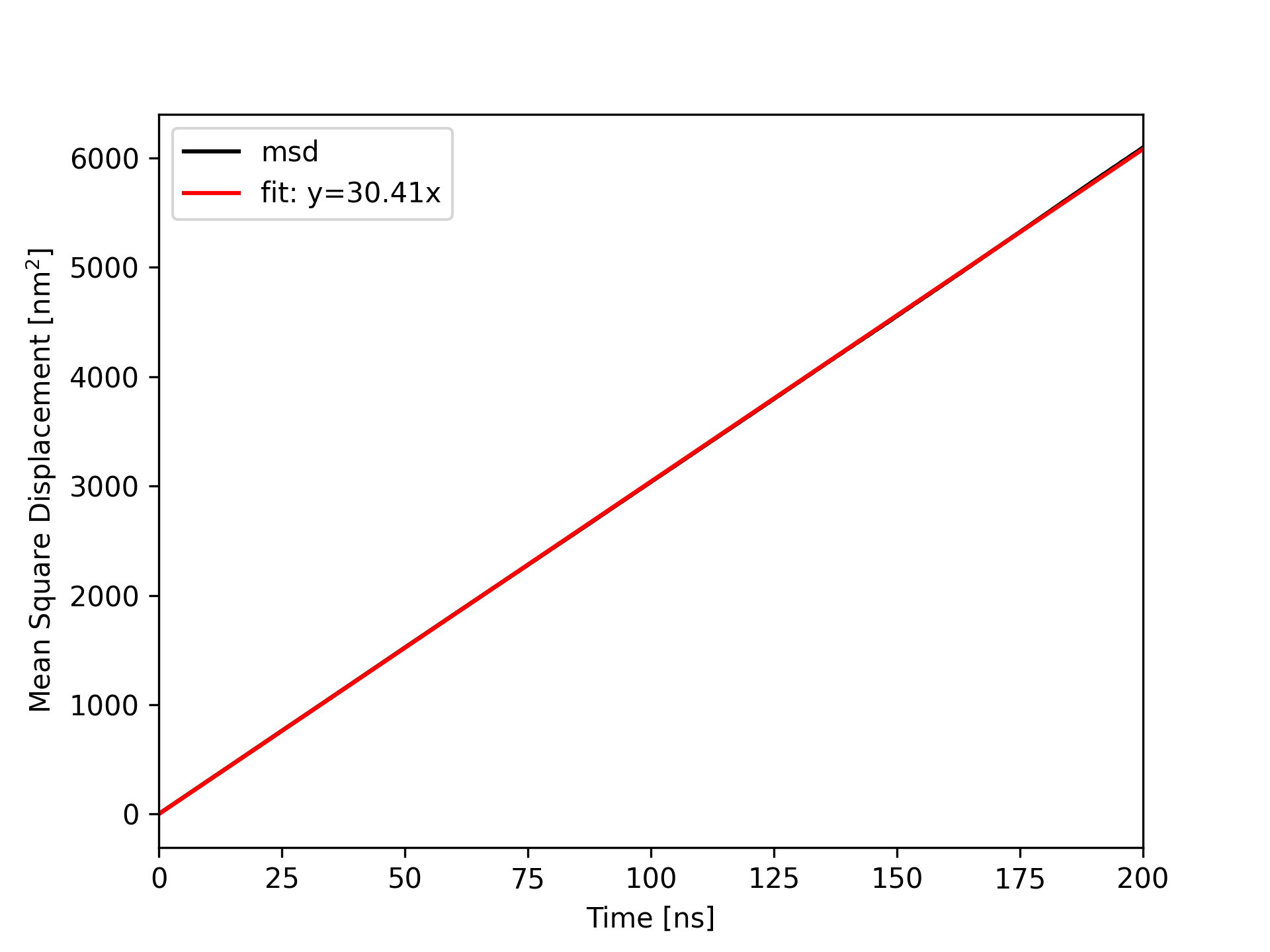}
   \caption{Mean square displacements for the NPT simulation in NAMD with the displacement method and a trajectory frequency of \SI{0.2}{\pico\second}.}
\end{figure*}
\begin{figure*}[ht]
	\centering
	\includegraphics[width=\linewidth]{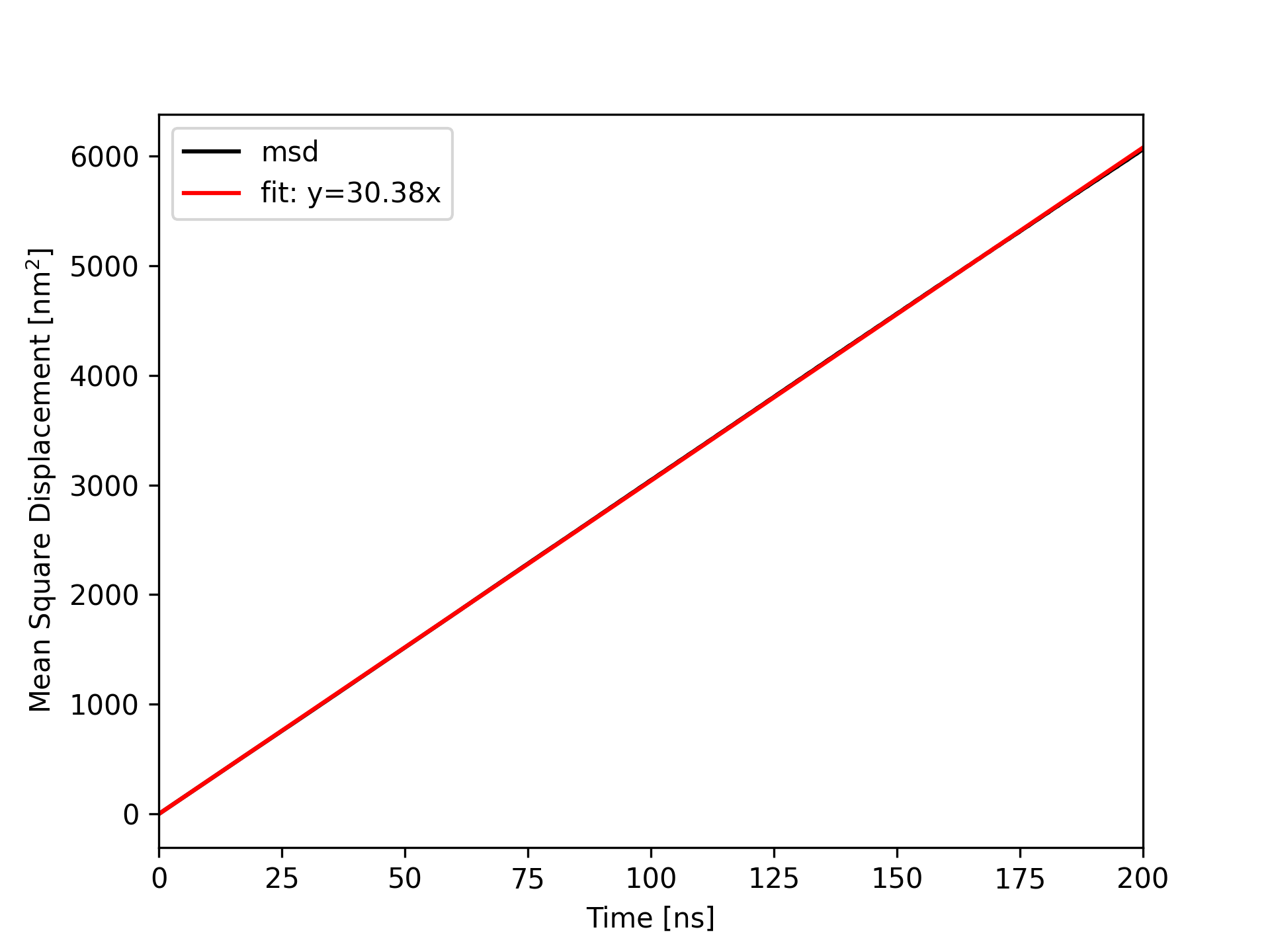}
   \caption{Mean square displacements for the NVT simulation in NAMD with the displacement method and a trajectory frequency of \SI{0.2}{\pico\second}.}
\end{figure*}
\begin{figure*}[ht]
	\centering
	\includegraphics[width=\linewidth]{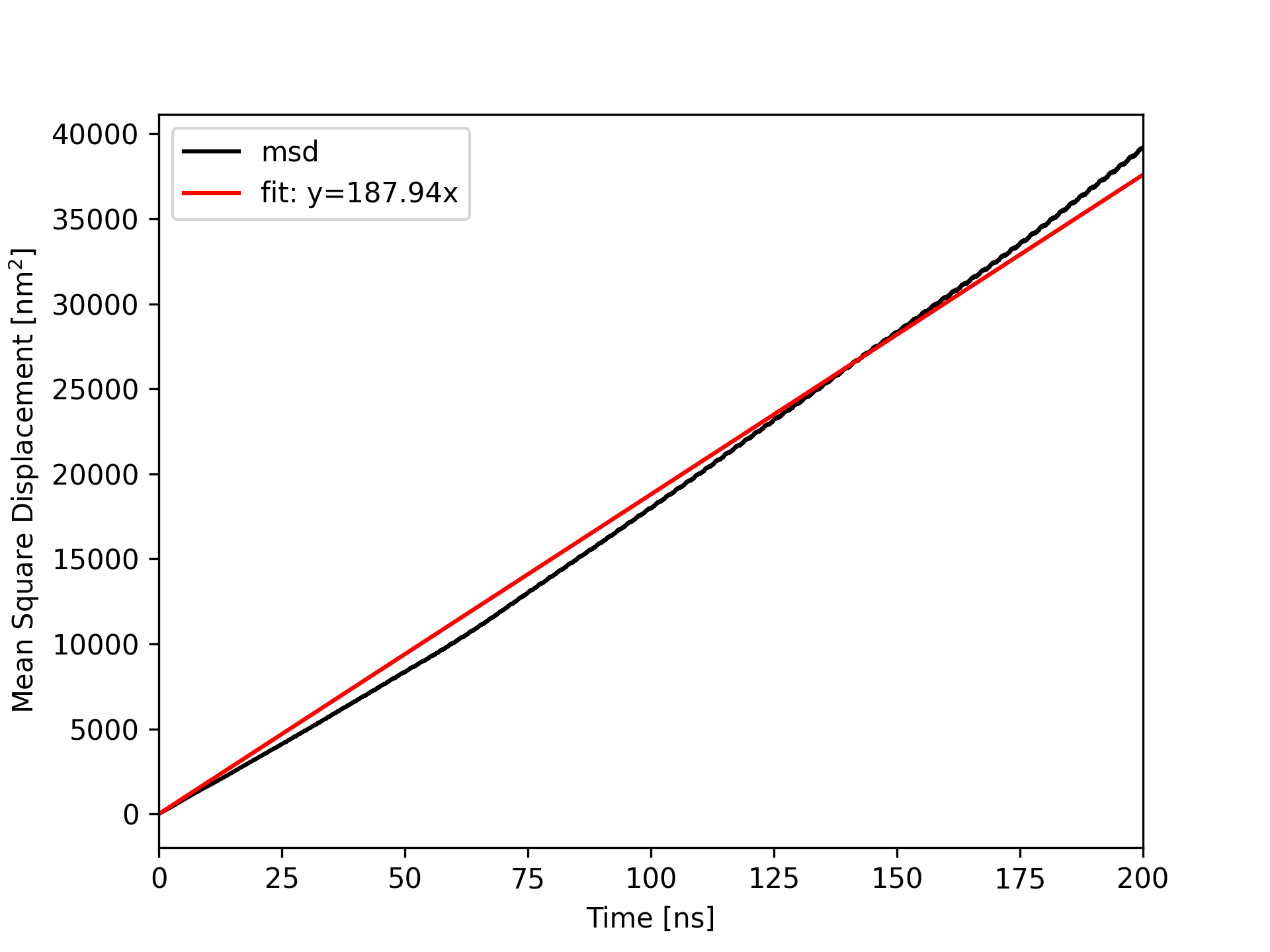}
   \caption{Mean square displacements for the NPT simulation in NAMD with the heuristic method and a trajectory frequency of \SI{2}{\pico\second}.}
\end{figure*}
\begin{figure*}[ht]
	\centering
	\includegraphics[width=\linewidth]{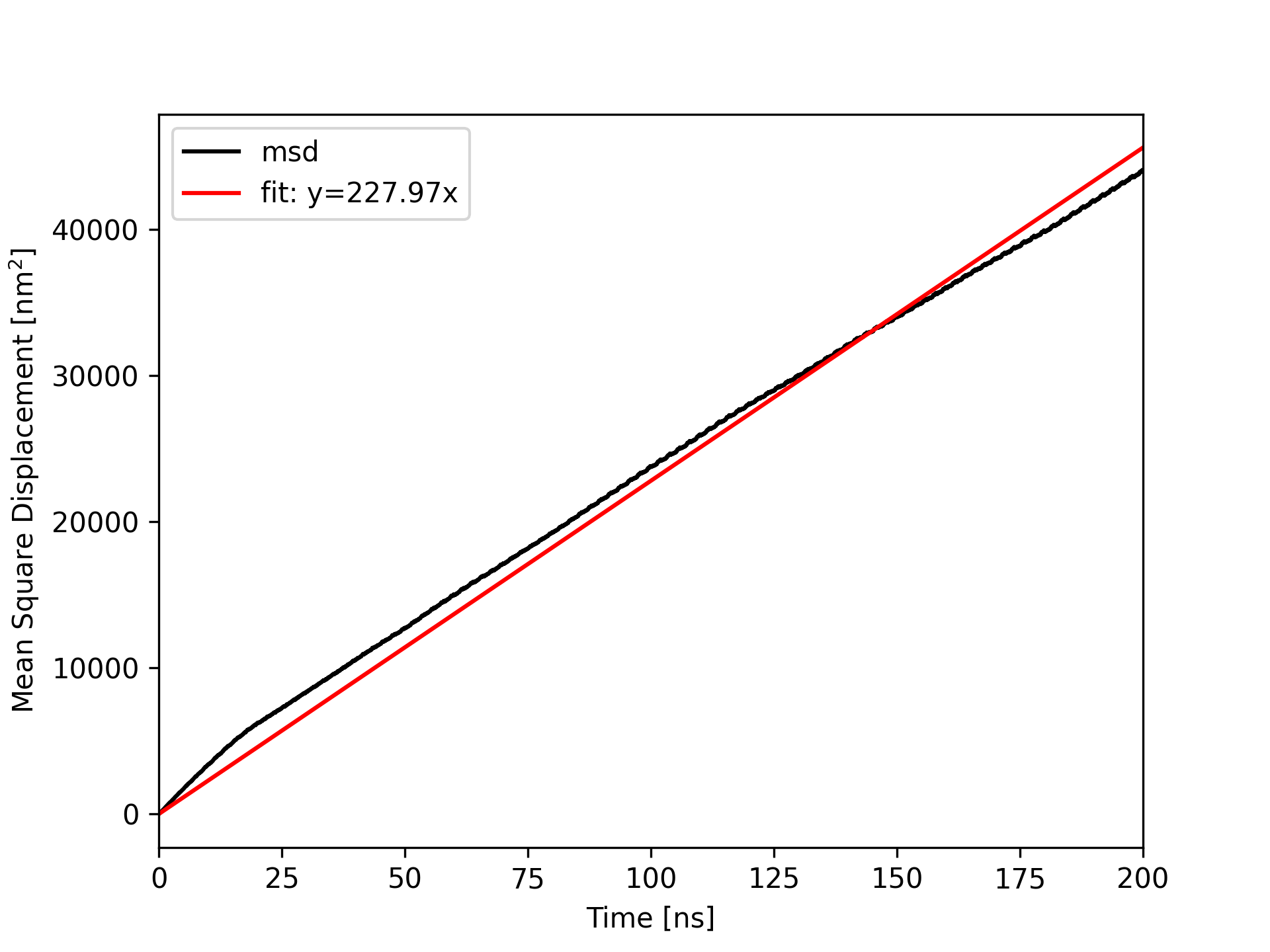}
   \caption{Mean square displacements for the NPT simulation in NAMD with the heuristic method and a trajectory frequency of \SI{1}{\pico\second}.}
\end{figure*}
\begin{figure*}[ht]
	\centering
	\includegraphics[width=\linewidth]{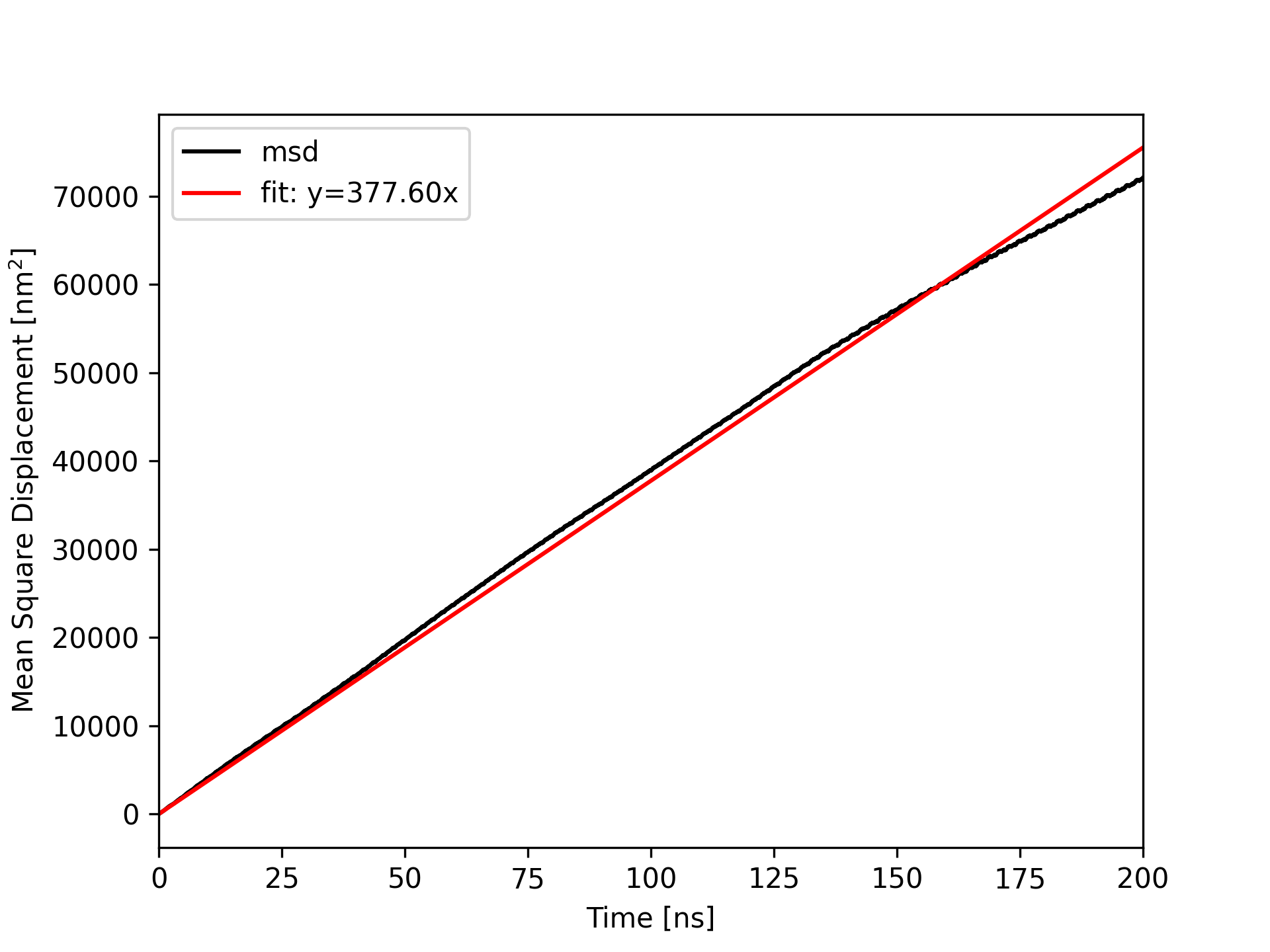}
   \caption{Mean square displacements for the NPT simulation in NAMD with the heuristic method and a trajectory frequency of \SI{0.5}{\pico\second}.}
\end{figure*}
\begin{figure*}[ht]
	\centering
	\includegraphics[width=\linewidth]{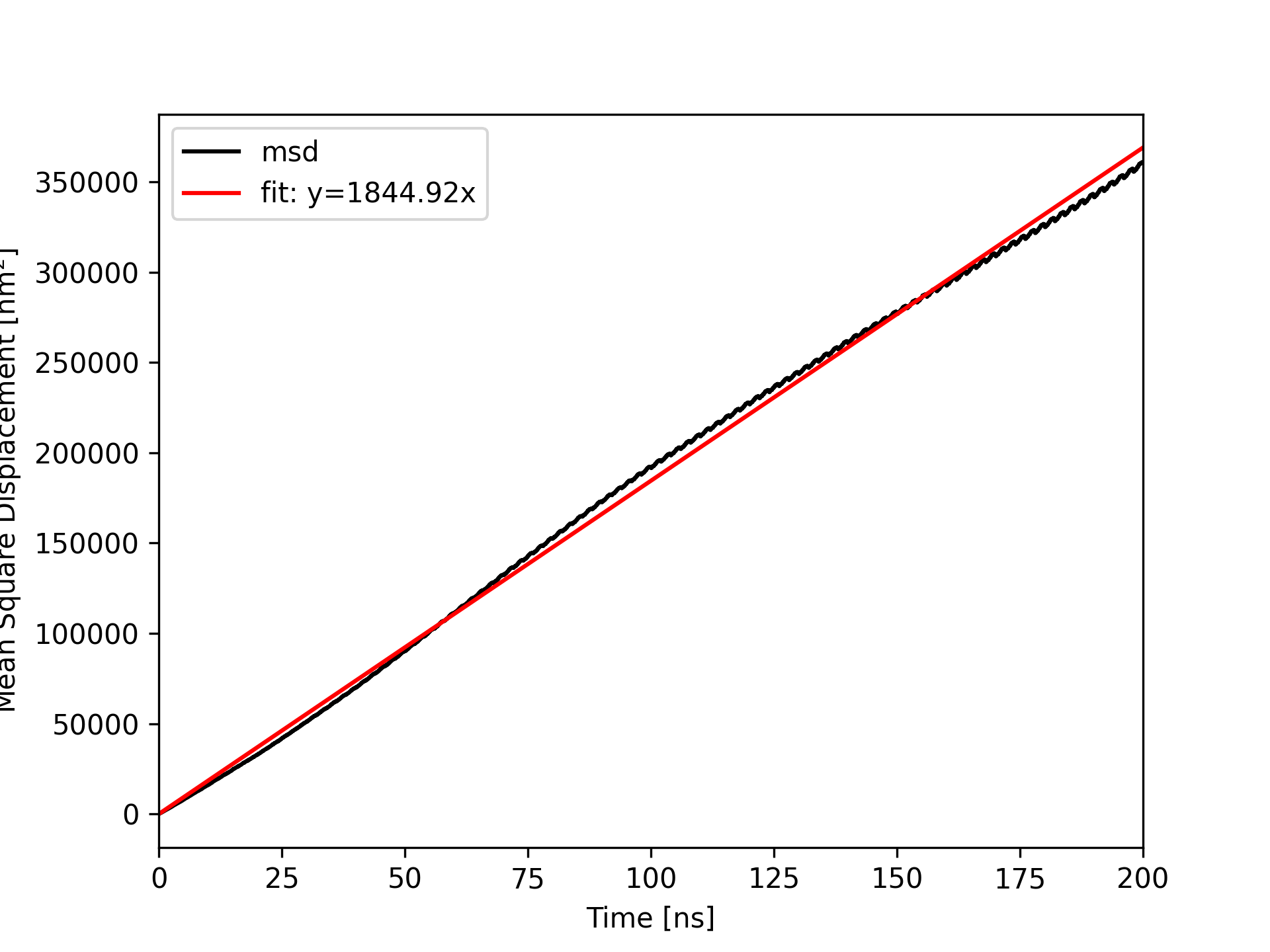}
   \caption{Mean square displacements for the NPT simulation in NAMD with the heuristic method and a trajectory frequency of \SI{0.2}{\pico\second}.}
\end{figure*}
\begin{figure*}[ht]
	\centering
	\includegraphics[width=\linewidth]{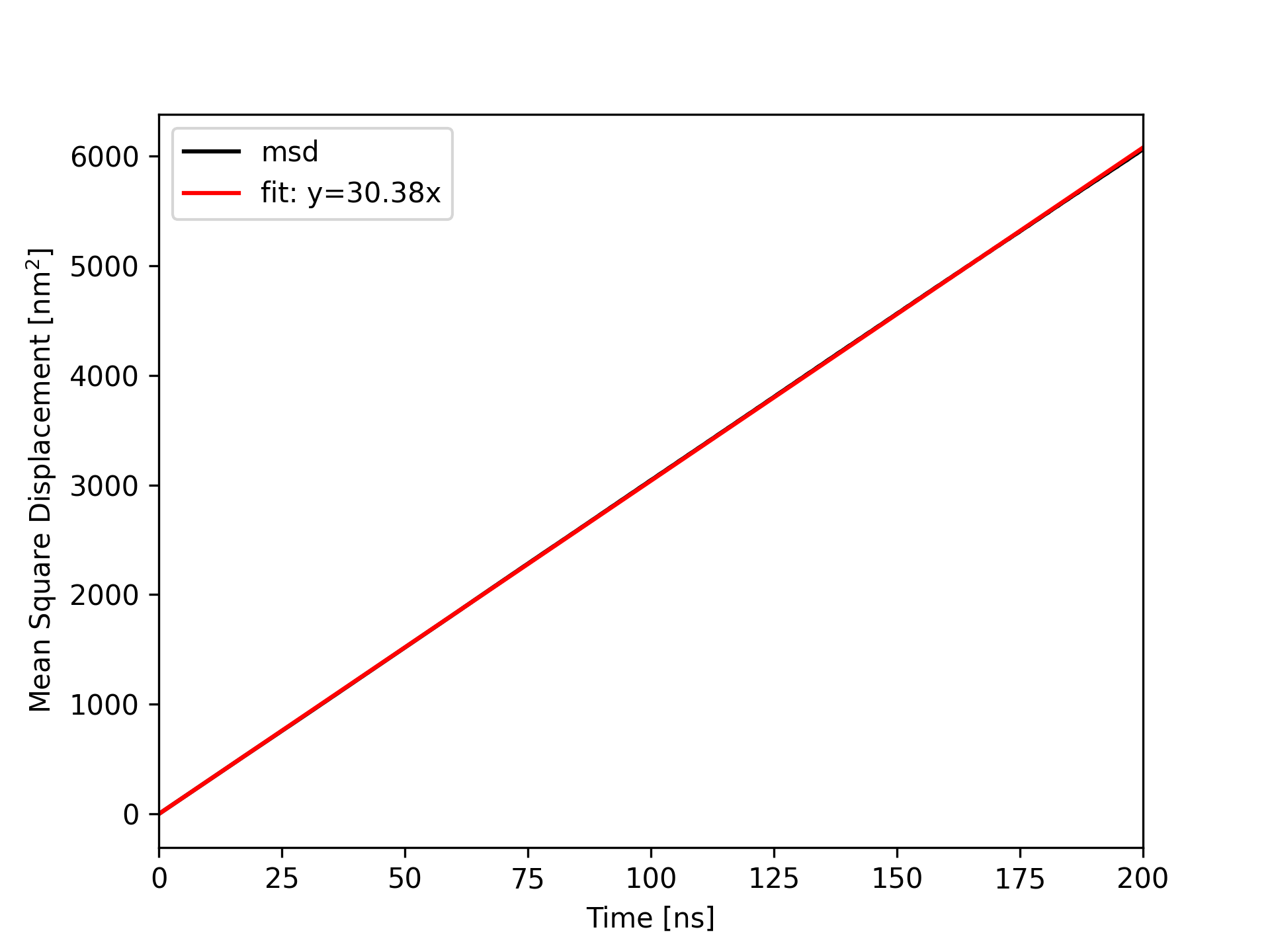}
   \caption{Mean square displacements for the NVT simulation in NAMD with the heuristic method and a trajectory frequency of \SI{0.2}{\pico\second}.}
\end{figure*}
\begin{figure*}[ht]
	\centering
	\includegraphics[width=\linewidth]{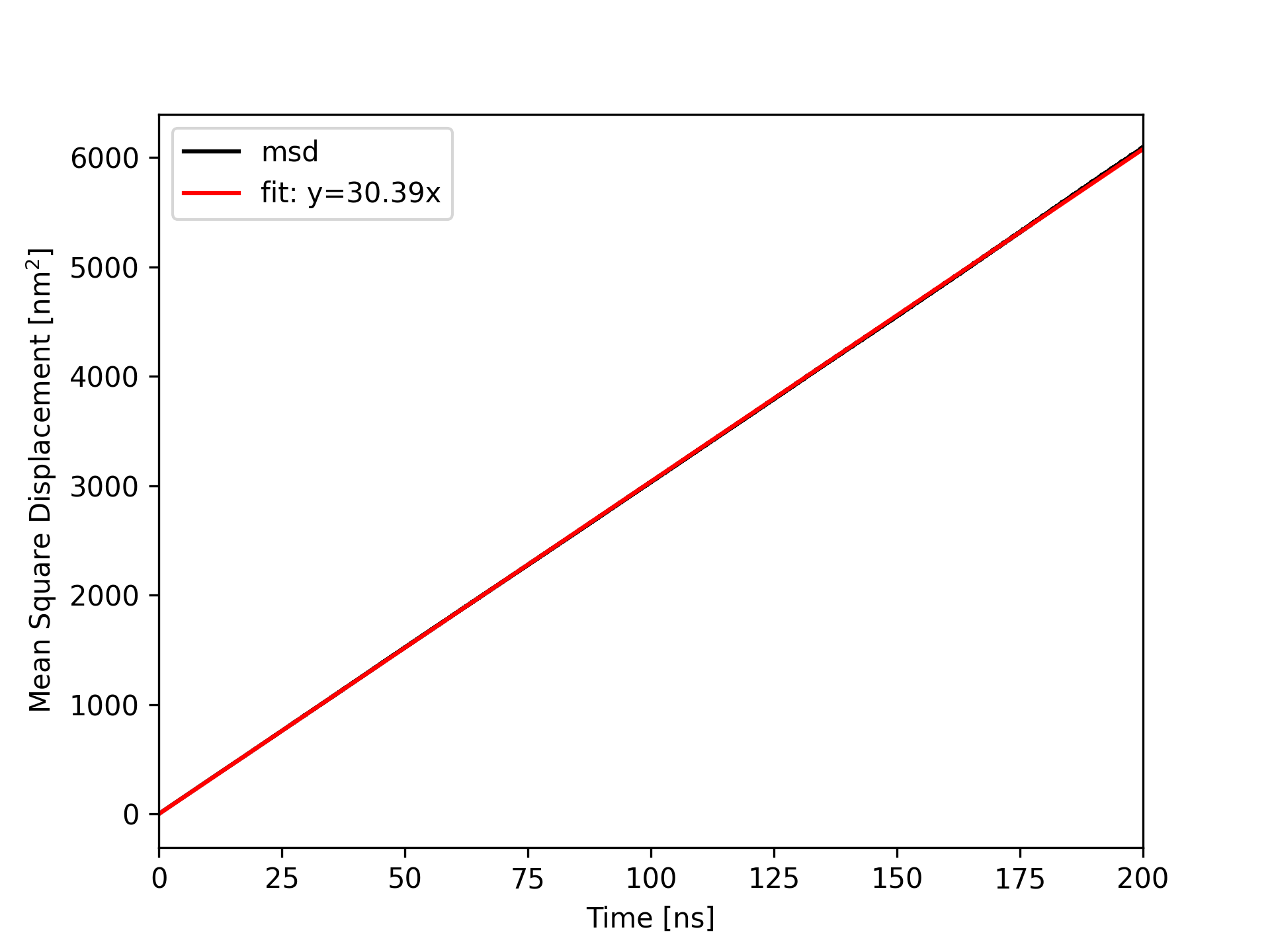}
   \caption{Mean square displacements for the NPT simulation in NAMD with the hybrid method and a trajectory frequency of \SI{2}{\pico\second}.}
\end{figure*}
\begin{figure*}[ht]
	\centering
	\includegraphics[width=\linewidth]{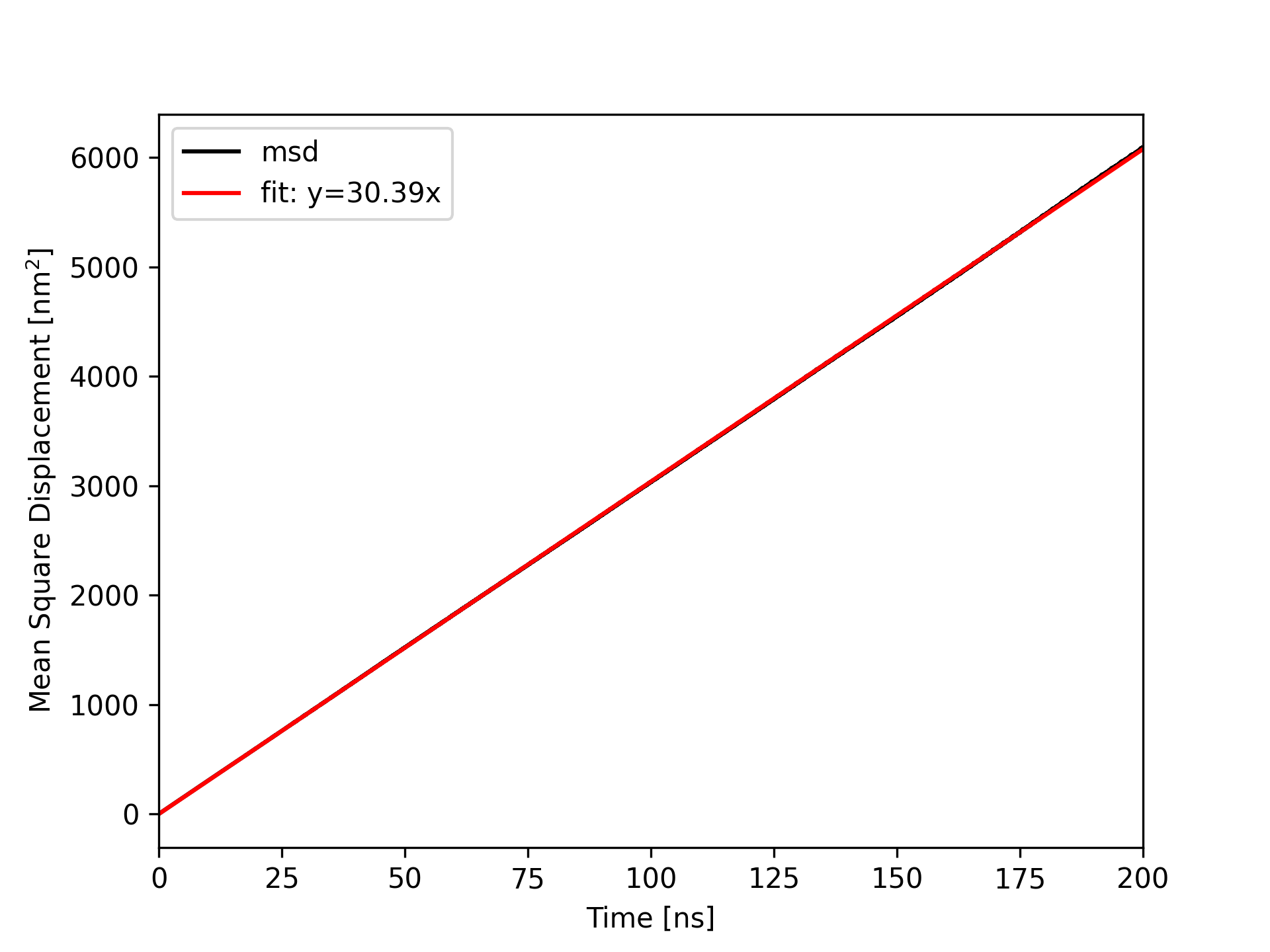}
   \caption{Mean square displacements for the NPT simulation in NAMD with the hybrid method and a trajectory frequency of \SI{1}{\pico\second}.}
\end{figure*}
\begin{figure*}[ht]
	\centering
	\includegraphics[width=\linewidth]{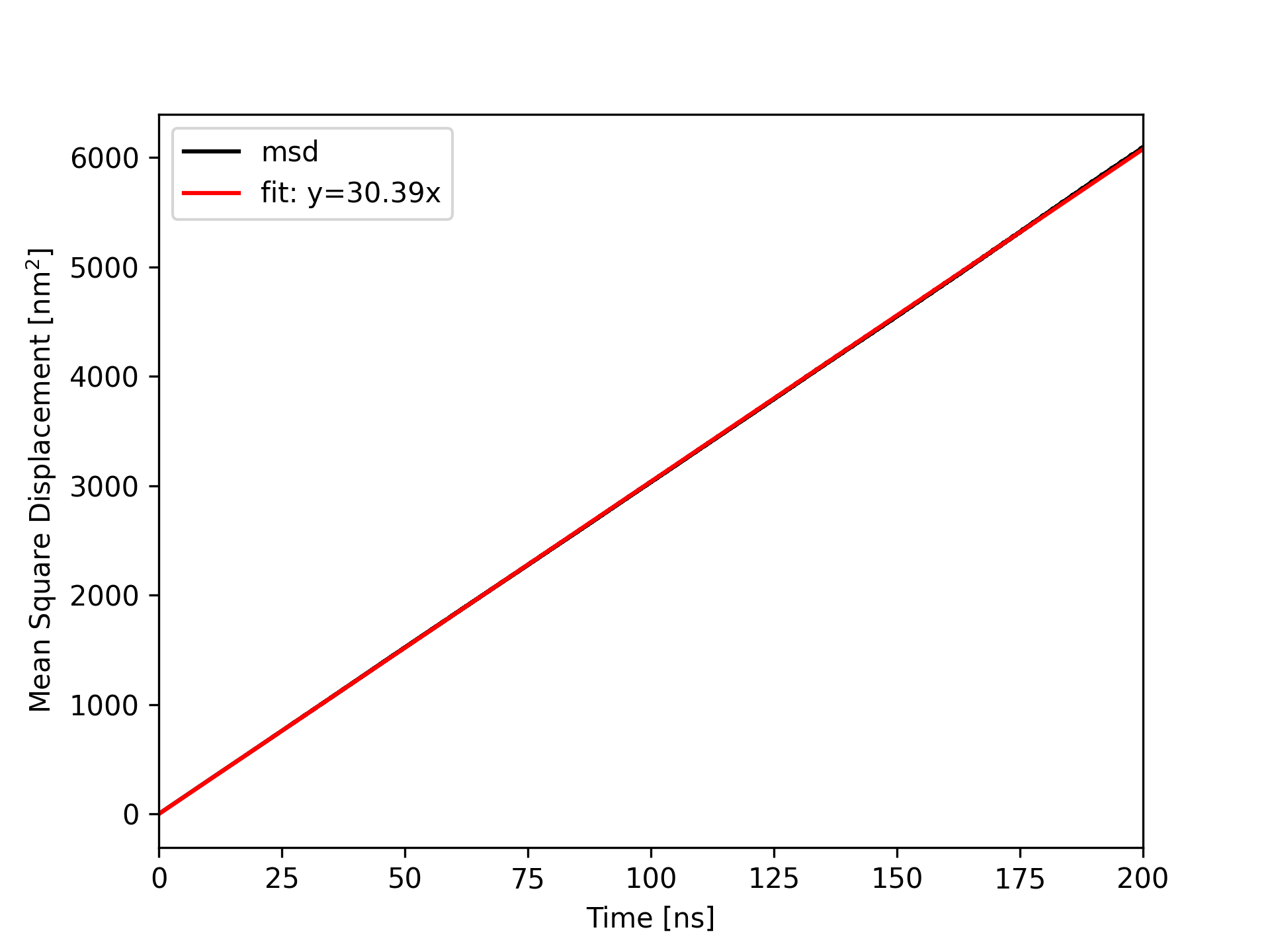}
   \caption{Mean square displacements for the NPT simulation in NAMD with the hybrid method and a trajectory frequency of \SI{0.5}{\pico\second}.}
\end{figure*}
\begin{figure*}[ht]
	\centering
	\includegraphics[width=\linewidth]{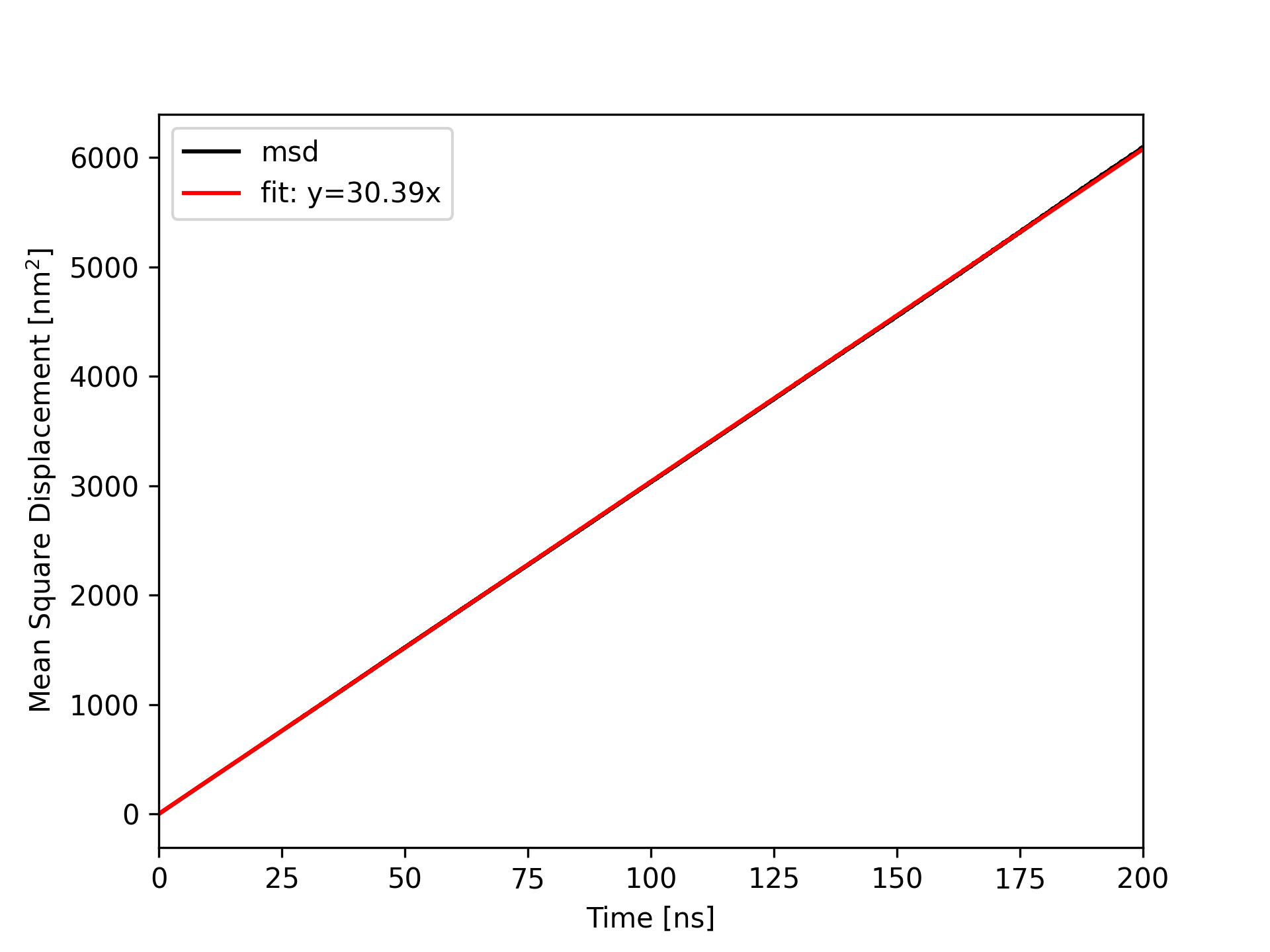}
   \caption{Mean square displacements for the NPT simulation in NAMD with the hybrid method and a trajectory frequency of \SI{0.2}{\pico\second}.}
\end{figure*}
\begin{figure*}[ht]
	\centering
	\includegraphics[width=\linewidth]{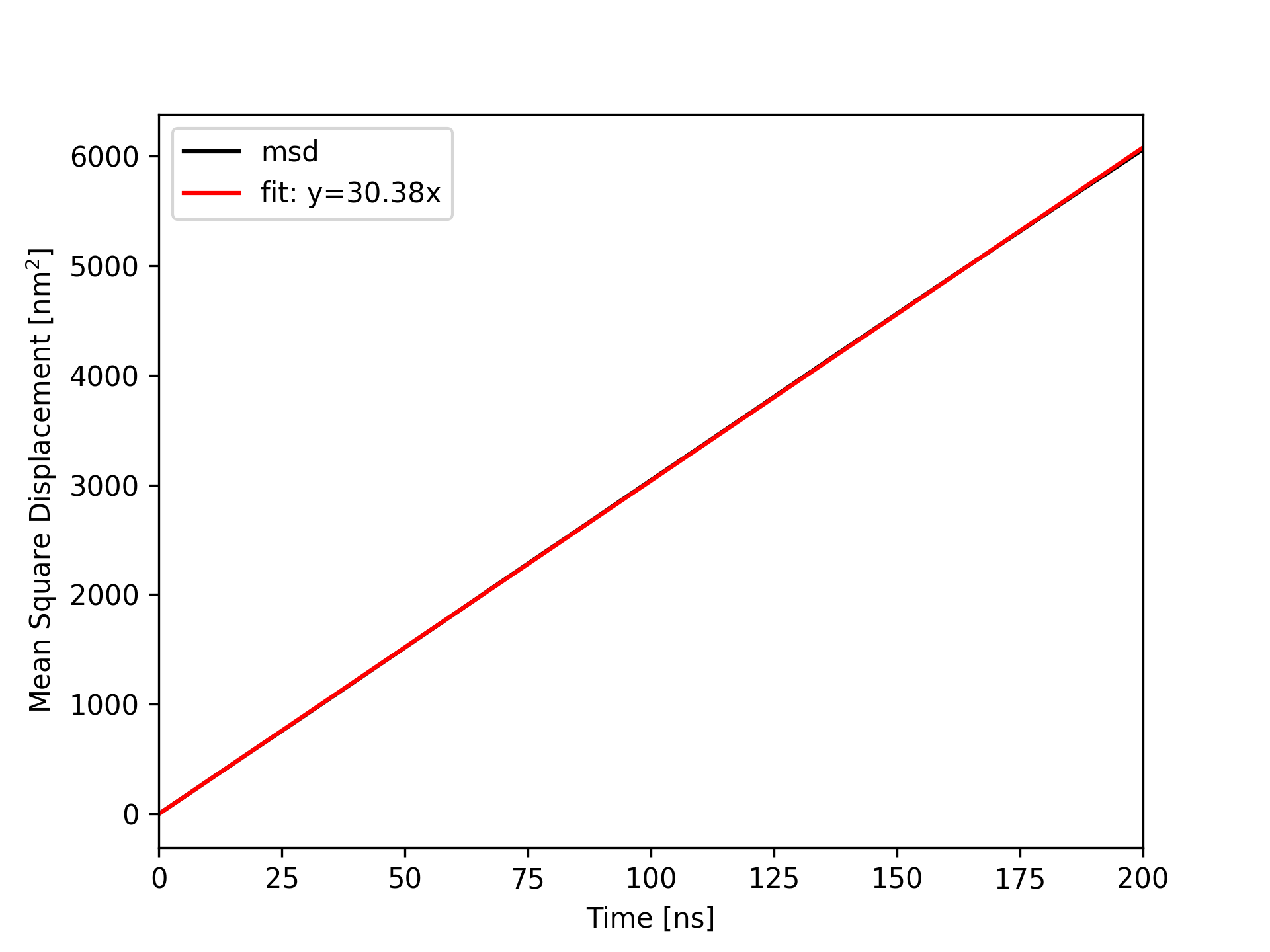}
   \caption{Mean square displacements for the NVT simulation in NAMD with the hybrid method and a trajectory frequency of \SI{0.2}{\pico\second}.}
\end{figure*}
\begin{figure*}[ht]
	\centering
	\includegraphics[width=\linewidth]{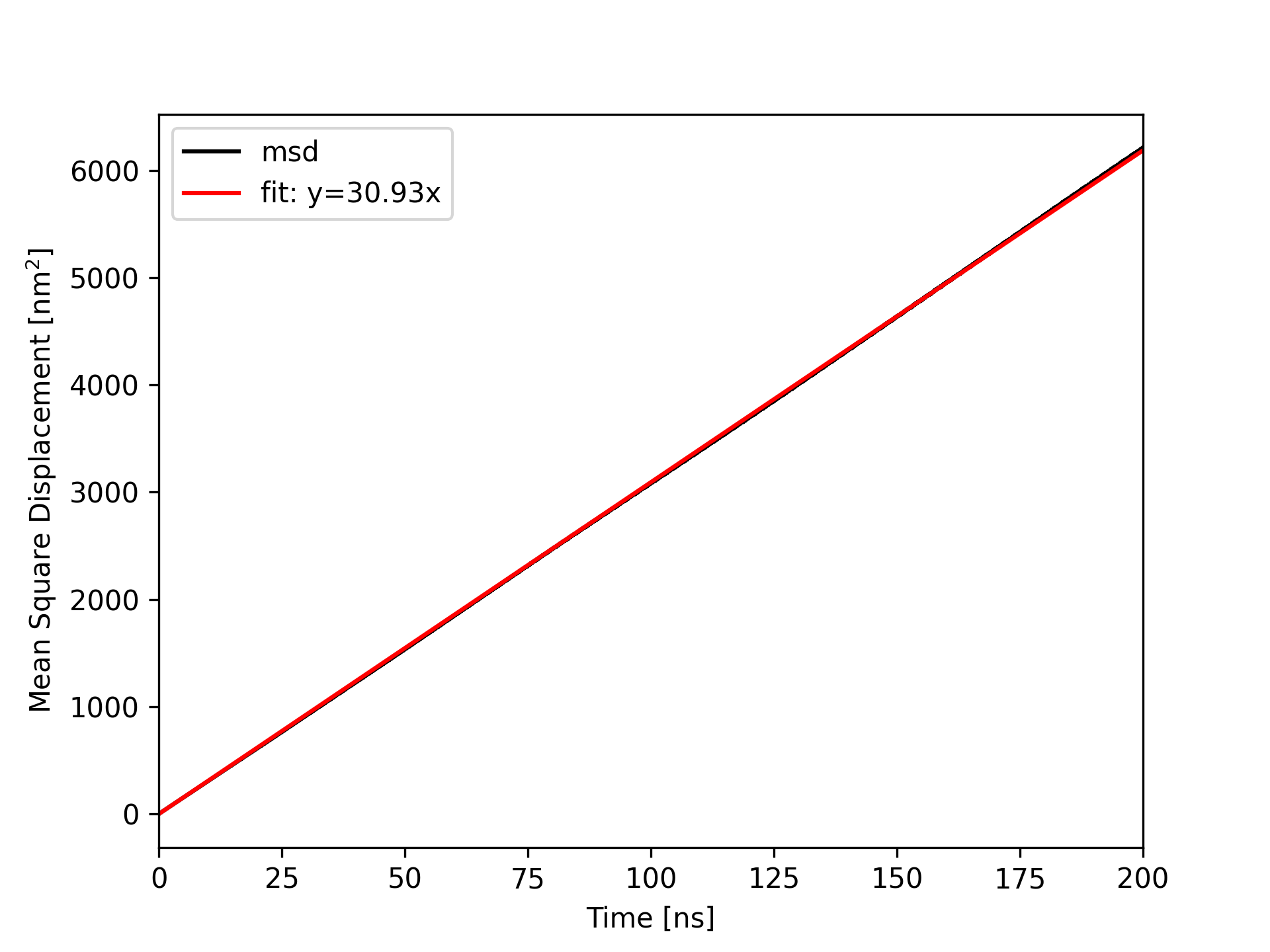}
   \caption{Mean square displacements for the reference simulation in NAMD with a trajectory frequency of \SI{0.2}{\pico\second}.}
\end{figure*}

\begin{figure*}[ht]
	\centering
	\includegraphics[width=\linewidth]{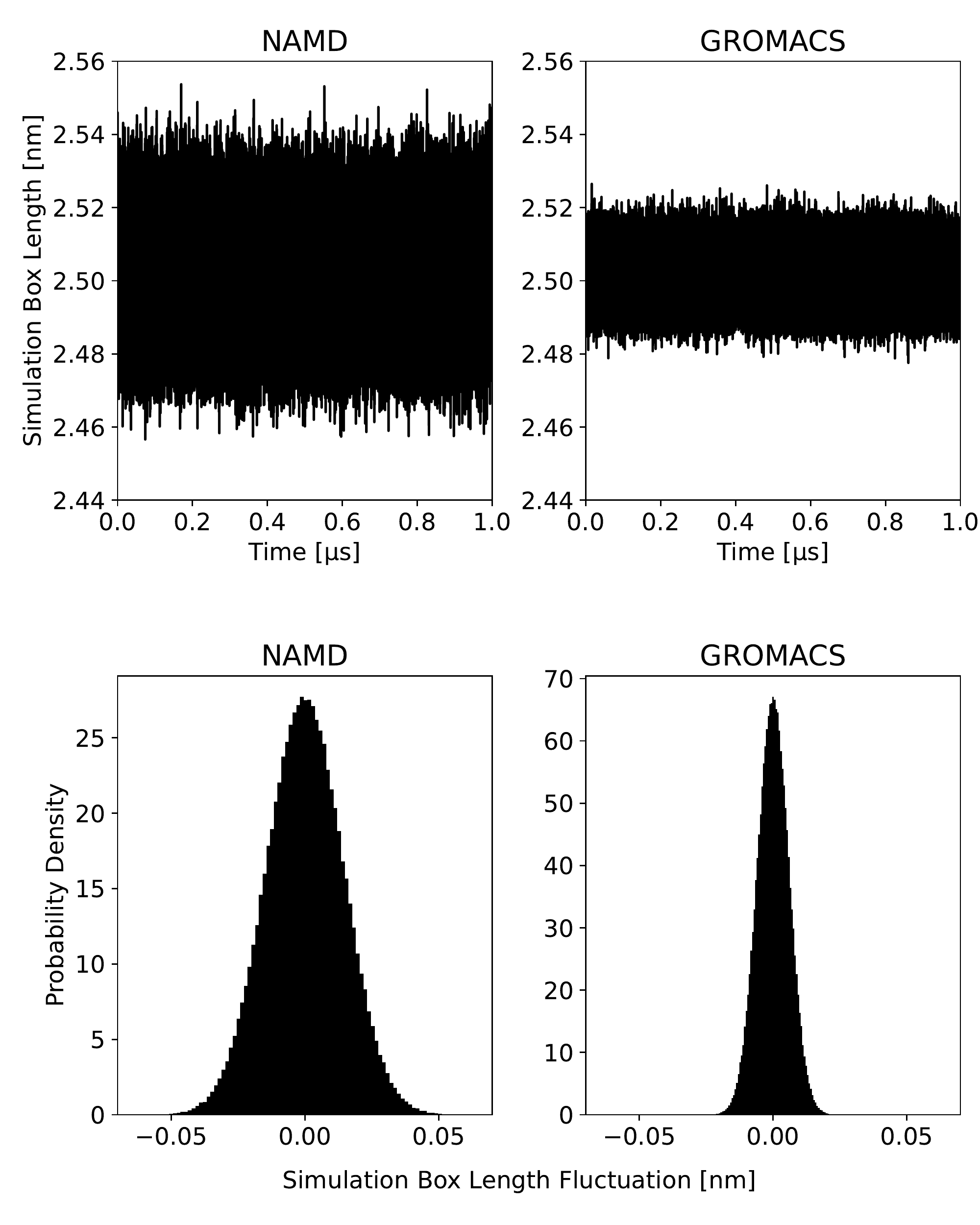}
	\caption{(Top) Isotropic box length fluctuations during the NPT simulations for NAMD (left) and GROMACS (right) over the simulation time. (Bottom) Histograms for the box length change after 2 ps intervals.
	The box volume fluctuates significantly more during the NAMD simulations.}
	\label{fig:si_fluctuations}
\end{figure*}

\begin{figure*}[ht]
	\centering
	\includegraphics[width=\linewidth]{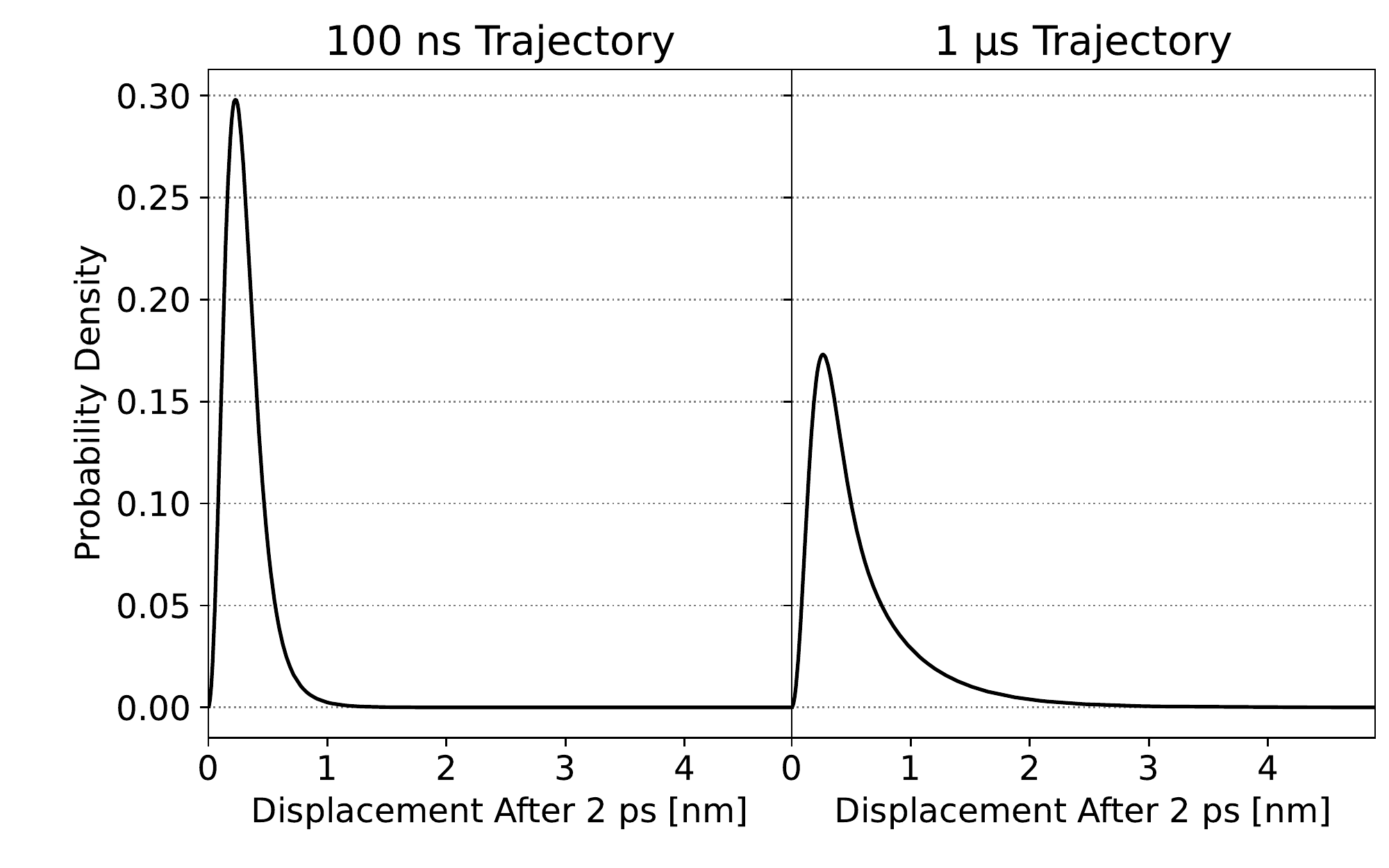}
	\caption{Histogram of the displacements after 2 ps intervals in the NAMD NPT reference simulation with unwrapped snapshots.
	The average displacement is around 0.25 nm, although rarely much higher displacements are possible.
    With increasing simulation time the molecules diffuse further away from the original simulation box.
	Due to the coordinate scaling of the barostat, the histograms get distorted with increasing simulation time, as molecules further away from the original simulation box are more effected by this scaling (compare left and right histogram).
    We cannot exclude that atoms moved more then half a box length during our simulation, resulting in an unwrapping error, but we assume that these events happen far less frequent then the other errors discussed in this manuscript.}
	\label{fig:si_disphist}
\end{figure*}